\documentclass[fleqn,usenatbib]{mnras}
\usepackage[T1]{fontenc}
\usepackage{verbatim}
\usepackage{newtxtext,newtxmath}

\DeclareRobustCommand{\VAN}[3]{#2}
\let\VANthebibliography\thebibliography
\def\thebibliography{\DeclareRobustCommand{\VAN}[3]{##3}\VANthebibliography}

\usepackage{graphicx}
\usepackage{amsmath}
  
\usepackage{bm}
\usepackage{subfig}
\usepackage{afterpage}
\usepackage{mathtools} 
\usepackage{float}
\usepackage{times}
\usepackage{color}
\usepackage{amsfonts}
\usepackage{booktabs}
\usepackage{siunitx}
\usepackage{soul}
\usepackage{xcolor}
\usepackage{float}
\usepackage{capt-of}
\usepackage{multicol}
\usepackage{multirow,multicol}
\usepackage{placeins}
\usepackage[normalem]{ulem}

\newcommand{\ie}{i.e.,~}

\title[Testing black hole space-times with the S2 star]{Testing black hole space-times with the S2 star orbit: a Bayesian comparison}

\author[Navarrete et. al.]
{C\'esar Navarrete$^{1}$,
Fernando V\'azquez-Ch\'avez$^{1,2}$,
Alejandro Cruz-Osorio$^{3\S}$, and
N\'estor Ortiz$^{1*}$\\
$^{1}$Instituto de Ciencias Nucleares, Universidad Nacional Aut\'onoma de M\'exico, AP 70-264, Ciudad de M\'exico 04510, M\'exico\\
$^{2}$DAMTP, Centre for Mathematical Sciences, University of Cambridge, Wilberforce Road, Cambridge CB3 0WA, UK\\
$^{3}$Instituto de Astronom\'{\i}a, Universidad Nacional Aut\'onoma de M\'exico, AP 70-264, Ciudad de M\'exico 04510, M\'exico \\ \\
$^{\S}$ aosorio@astro.unam.mx\\
$^{*}$ nestor.ortiz@nucleares.unam.mx
}

\usepackage{hyperref}
\hypersetup{
  colorlinks=true,        
  linkcolor=blue,         
  citecolor=cyan,         
  filecolor=magenta,      
  urlcolor=magenta        
}

\begin{document}
\label{firstpage}
\pagerange{\pageref{firstpage}--\pageref{lastpage}}
\maketitle

\begin{abstract}
We implement a Markov Chain Monte Carlo method to obtain posterior probability distributions for the parameters of the S2 star orbit around Sagittarius\,A*, for seven representative non-rotating black hole space-time solutions. In particular, we consider the Schwarzschild, Reissner-Nordstr\"om, Janis-Newman-Winicour, and Bardeen black hole space-times from General Relativity, as well as a black hole solution from Einstein-Maxwell-dilaton gravity, a hairy black hole solution from Horndeski theory, and a Yukawa-like black hole from $f(\mathcal R)$ gravity. To constrain model parameters, we use the most recent publicly available observational data of the S2 star orbit, namely astrometric measurements, spectroscopic data, and the pericentre advance measured by the GRAVITY Collaboration. We further perform a consistent Bayesian comparison of models, calculating the log-Bayes factor of each space-time with respect to the Schwarzschild solution.
Our results show that the currently available data indicate no statistically significant preference among the space-times considered. The Bardeen and Yukawa-like models are indistinguishable from Schwarzschild within current uncertainties, while the Reissner-Nordstr\"om, Janis-Newman-Winicour, Horndeski and Einstein-Maxwell-dilaton geometries show at most weak and non-decisive strength of evidence under the adopted priors and likelihood choices.
\end{abstract}

\begin{keywords}
black hole physics -- methods: data analysis -- astrometry -- Galaxy: centre.
\end{keywords}

\section{Introduction} 
\label{sec:intro}
In 1974, astronomers identified a bright radio source in the direction of the Sagittarius constellation at the heart of the Milky Way, which became known as Sagittarius A* (Sgr\,A*) \citep{DescSgrA*}. Follow-up observations of the motion of nearby gas clouds revealed an extraordinarily compact and massive object at the heart of our Galaxy---the region now referred to as the Galactic Centre \citep[for a review see][]{Reid:2008fp}.

The motion of Sgr\,A* with respect to a background quasar, namely J1745-283, was found to be consistent with that of a stationary object located at the dynamical centre of the Milky Way \citep{2004ApJ...616..872R}. However, by the end of the 20th century, the possibility that Sgr\,A* was simply a dense cluster of stars had not yet been ruled out. Because optical and ultraviolet light are absorbed by interstellar dust, direct observations of the Galactic Centre remained challenging. Astronomers then turned to infrared and other wavelengths to study the motion of stars in the vicinity of Sgr\,A* in order to probe its nature more directly.

Recently, the Event Horizon Telescope (EHT) Collaboration captured the first image of Sgr\,A* through Very Long Baseline Interferometry, finding strong evidence in favor of the engine of Sgr\,A* being a Supermassive Black Hole (SMBH) surrounded by an accretion disk of hot plasma \citep{EHT}. The measured shadow diameter and shape agree, within a 17\% margin \citep{Kocherlakota2021,Psaltis2020_EHT}, with predictions from the Kerr space-time---a rotating black hole in General Relativity (GR). However, measurement uncertainties still leave room for alternative black hole models. In this context, the EHT collaboration analysis compares the observed image with predictions from both Kerr and non-Kerr space-times, placing constraints on possible deviations from GR.

Over the past two decades, astronomers have tracked the orbits of stars around the Galactic Centre with remarkable precision \citep{RevModPhys.82.3121}. These observations reveal that the stars follow nearly perfect elliptical paths around a central mass confined within a radius of roughly $100$ AU \citep{Reid:2008fp}, with acceleration vectors pointing directly toward a gravitational source located extremely close to Sgr\,A*. Such motion provides compelling evidence that the heart of our Galaxy hosts an SMBH \citep{ghez1998high, ghez2008measuring}. Beyond confirming the SMBHs presence, the accumulated data offer a unique laboratory for testing gravity theories---\citep[see, e.g., ][and references therein]{Della_Monica_2022}. Among these stellar probes, the star S2 stands out as a particularly valuable test case. With a pericentre of $\approx 120\, {\rm AU}$ and a peak orbital velocity of $\approx 7,700$ km/s ($\sim 2.5\%$ of the speed of light), S2 ventures deep enough into Sgr\,A*'s gravitational well to exhibit clear relativistic effects. Indeed, both its gravitational redshift \citep{S2Redshift,abuter2018detection} and orbital precession \citep{abuter2020detection} have been detected and analyzed within the framework of GR.

Previous studies on the S2 star orbit have placed constraints on the parameters of various space-time solutions, including Schwarzschild \citep{S2Redshift}, a Schwarzschild-like metric of Scalar-Tensor-Vector Gravity \citep{della2022orbital}, the spherically symmetric parameterized Rezzolla-Zhidenko (PRZ) space-time \citep{shaymatov2023parameters}, the weak-field limit of $f(\mathcal R)$-gravity \citep{de2021f}, a subclass of asymptotically flat Buchdahl-inspired vacuum space-times \citep{yan2024observational}, Einstein-Maxwell-dilaton-axion gravity \citep{fernandez2023constraining}, a non-linear Yukawa-like correction to the Newtonian gravitational potential \citep{jovanovic2024constraints}, the weak-field limit of Horndeski theory \citep{della2023testing}, the q-metric naked singularity \citep{lora2023q}, and the Janis-Newman-Winicour \citep{bambhaniya2024relativistic}  space-times. These works have employed diverse methods and assumptions; for instance, orbital precession has been computed under varying approximations---or neglected at all. As a consequence, the results are not directly comparable and the data preference for each model cannot be meaningfully assessed.

In this work, we examine seven representative black hole solutions from GR and beyond, including several space-times studied in previous works. For the first time, we model the S2 star orbit consistently across all cases using unique prescriptions to compute orbital precession, R{\o}mer delay, redshift effects, initial conditions, and related quantities using the most recent publicly available data of the S2 star orbit. Other novelties of this work include $(i)$ the first analysis of the Bardeen regular black hole solution using the S2 star orbit; $(ii)$ a more robust analysis of the Janis-Newman-Winicour solution, including the precession of S2 and not excluding the possibility of a vanishing scalar charge; and $(iii)$ we do not assume the approximation $\Psi \sim \Phi$ for the Yukawa-like space-time---see Section~\ref{sec:Yukawa} for details.

Assuming uniform prior distributions for the orbital parameters, we apply a Markov Chain Monte Carlo (MCMC) algorithm to obtain their posterior distributions and corresponding confidence intervals. Such a consistent framework allows for a direct comparison of posterior distributions (for shared parameters) and best-fit values across different models. Furthermore, to quantify the data preference---given our chosen priors---we perform a Bayesian comparison of the space-time models under consideration.

The structure of the paper is as follows. In Section~\ref{sec:setup}, we present the general features of our representative space-time catalog, and the prescription used to compute orbital precession. Section~\ref{sec:MCMC} describes the orbit modelling, the implementation of MCMC methods to obtain posterior distributions of orbital parameters, and relevant effects in the trajectories. Section~\ref{sec:results} discusses key aspects of the posterior distributions, provides bounds for the parameters of each space-time, and presents the Bayesian comparison, including the minimum $\chi^2$ and reduced $\chi_\nu^2$ values for each case. Section~\ref{sec:conclusion} summarizes our findings.

\section{Non-rotating space-time catalogue} \label{sec:spacetimes}
\label{sec:setup}
The space-time catalog considered in this work is not exhaustive but rather a selection of representative solutions of non-rotating black holes, including naked singularities, charged, regular, and hairy black holes. Several space-times are excluded---not because they are irrelevant, but due to either known inconsistencies with observations or the lack of predictions for stellar orbits compared to other models. 

Among the excluded space-times are the Joshi-Malafarina-Narayan naked singularity space-times (JMN-1 and JMN-2) \citep{Joshi_2011, Joshi_2014}. The JMN-1 metric is indistinguishable from Schwarzschild solution beyond a radius $R_{b}$, constrained to $R_{b} \leq 3M$ for the space-time to possess a photon sphere and cast a shadow \citep{Shaikh2018ShadowsOS}. In the case of Sgr\,A*, the S2 star orbit is beyond $R_b$, meaning that this model would produce no observable effect on the star geodesic motion. We therefore exclude JMN-1 from our analysis. In turn, JMN-2 has been recently ruled out by the EHT Collaboration \citep{EHT} due to the absence of a photon sphere for any physically allowed parameters, and thus failure to predict the formation of a shadow.

The line element describing non-rotating black holes---i.e., spherically symmetric, static space-times---is given by 

\begin{eqnarray}\label{eq:generic_line_element}
\mathrm{d} s^2=-f(r) \mathrm{d} t^2+g(r) \mathrm{d} r^2+h(r) \mathrm{~d} \Omega^2,
\end{eqnarray}
written in Schwarzschild-like coordinates $(t,r,\theta,\phi)$, where $\mathrm{d}\Omega^2=\mathrm{d} \theta^2+\sin ^2 \theta \mathrm{d} \phi^2$ denotes the line element on the two-sphere. Each space-time considered here corresponds to a specific choice of the functions ${f, g, h}$. Throughout this paper, we adopt the metric signature $(-, +, +, +)$. Additionally, unless otherwise specified, we use geometrized units, such that $c = G = 1$.
In the following, we describe the seven black holes used to model the space-time near Sgr\,A*.

\subsection{Schwarzschild black hole}
The Schwarzschild metric \citep{Schwarzschild} is the exact solution of the Einstein field equations describing the space-time outside a spherically symmetric black hole of mass $M$. For this case, the functions corresponding to the line element in Eq. \eqref{eq:generic_line_element} are
\begin{eqnarray}
f(r)=g(r)^{-1}=1-\frac{r_s}{r}, \qquad h(r)=r^2 ,
\end{eqnarray}
where $r_s=2M$ denotes the Schwarzschild radius, which coincides with the black hole's event horizon.

\subsection{Reissner-Nordstr\"om charged black hole}
The Reissner-Nordstr\"om (RN) metric 
\citep{Reissner, Nordstrom} describes the space-time around a spherically symmetric, electrically charged black hole of mass $M$. The components of the line element \eqref{eq:generic_line_element} for this space-time are given by
\begin{eqnarray}
f(r)=g(r)^{-1}=1-\frac{2M}{r}+\frac{Q^{2}}{r^{2}}, \qquad h(r)=r^2,
\end{eqnarray}
where $Q$ is the black hole's electric charge. The Schwarzschild solution is recovered when $Q=0$. In addition to an event horizon, the RN black hole has an internal Cauchy horizon. Both of them coincide when the black hole becomes extremal, i.e. when $2Q = r_s$.

\subsection{Janis-Newman-Winicour black hole}
The Janis-Newman-Winicour (JNW) space-time is a spherical solution to the Einstein-scalar field system, sourced by a real, minimally coupled, massless scalar field \citep{JNW}. It describes a static naked singularity. Following the notation of \cite{Virbhadra_1997}, the components of the line element \eqref{eq:generic_line_element} are
\begin{eqnarray}
f(r)=g(r)^{-1}=\left ( 1-\frac{b_{_\text{JNW}}}{r} \right )^{\gamma}, \quad
h(r)=r^2 \left ( 1-\frac{b_{_\text{JNW}}}{r} \right )^{1-\gamma},
\end{eqnarray}
where $ b_{_\text{JNW}} = 2\sqrt{ M^{2}+q_{_\text{JNW}}^{2}}$, and $\gamma = 2 M/b_{_\text{JNW}}$. Here, $M$ is the mass parameter, while $q_{_\text{JNW}}$ represents the scalar charge of the space-time. In the case of $q_{_\text{JNW}} = 0$, the Schwarzschild solution is recovered.

\subsection{Bardeen regular black hole}
\label{sec:bhac}
The Bardeen space-time \citep{Bardeen1968} is a spherically symmetric solution of the Einstein equations, sourced by a self-gravitating magnetic field in nonlinear electrodynamics, characterized by its monopole charge $g$ \citep{Ayon}. This solution is regular in the sense that it avoids the formation of a central curvature singularity. The components of the corresponding line element \eqref{eq:generic_line_element} are
\begin{eqnarray}
f(r) =g(r)^{-1}=1-\frac{2M r^2}{\left(r^2+g^2\right)^{3 / 2}}, \qquad
h(r)=r^2,
\end{eqnarray}
with $M$ the Arnowitt-Deser-Misner (ADM) mass of the space-time. For $g^2 \le 16M^2/27$, the Bardeen metric admits an event horizon, in which case it describes a regular black hole that becomes extremal when $g^2 = 16M^2/27$. In the case of $g=0$ one recovers the Schwarzschild black hole. Remarkably, in this space-time, the orbital precession of a massive test body coincides, to first order, with that of the Schwarzschild solution \citep{li2023quasi}.

\subsection{Hairy black hole in Horndeski gravity}
The Horndeski theory of gravity \citep{Horndeski} is the most general scalar-tensor theory in four dimensions in which the Lagrangian yields second-order equations of motion. In particular, we consider the spherical, exact hairy black-hole solution obtained by \citet{bergliaffa2021hairy}, which is given by
\begin{eqnarray}
f(r) =g(r)^{-1}=1-\frac{2M}{r}+\frac{q_\text{H}}{r} \ln \left(\frac{r}{2M}\right), \quad
h(r)=r^2,
\end{eqnarray}
where $q_{\text{H}}$ is a scalar charge. The Schwarzschild solution is recovered when $q_{\text{H}}=0$. For practicality, throughout this paper we refer to this solution as \textit{Horndeski space-time}.

\subsection{Yukawa-like black hole from $f(\mathcal R)$ gravity}
\label{sec:Yukawa}
In $f(\mathcal R)$ theories of gravity, the Einstein-Hilbert action is generalized by replacing the Ricci scalar $\mathcal{R}$ by a general function $f(\mathcal{R})$. Expanding the function $f$ in a Taylor series, and keeping terms up to order $1/c^2$ in the field equations, yields spherically symmetric solutions with a Yukawa-like correction to Newton's gravitational potential in the weak-field limit \citep{ Yukawa2}. In this space-time, the functions of the line element \eqref{eq:generic_line_element} are
\begin{eqnarray}
f(r) =1+\Phi(r), \qquad
g(r)=1-\Psi(r), \qquad
h(r)=r^2, 
\end{eqnarray}
with the potentials $\Phi$ and $\Psi$ given by
\begin{eqnarray}
\Phi(r)&=&-\frac{2M}{r(\delta+1)}\left( 1 + \delta e^{-\frac{r}{\lambda}}\right)~, \\
\Psi(r)&=&-\frac{2M}{r(\delta+1)}\left[1- \left( 1 + \frac{r}{\lambda} \right) \delta e^{-\frac{r}{\lambda}}\right]~.
\label{eq:Psi}
\end{eqnarray}
Here, $M$ is the mass of the source, $\delta$ quantifies the strength of the Yukawa-like potential and deviations from GR, and $\lambda$ denotes a characteristic length scale. In the case of $\delta=0$, one recovers the weak-field limit of the Schwarzschild space-time. By varying $\delta$ and $\lambda$, one obtains black holes with different event-horizon and photon-ring sizes. For practicality, this solutions are referred to as \textit{Yukawa-like black holes}. For a detailed analysis, see \citet{Cruz2021}.

\subsection{Einstein-Maxwell-dilaton black hole}
The Einstein-Maxwell-dilaton (EMd) theory of gravity arises in the low-energy limit of string theory, where the dilatonic field couples linearly to the electromagnetic sector. The corresponding field equations admit a solution describing an electrically charged black hole \citep{EMd1Strominger, garcia1995class}. In this case, the line element functions are
\begin{eqnarray}
f(r)=g(r)^{-1}=1-\frac{2M}{r +  r_{2}}, \qquad
h(r)=r\left ( r+r_{2} \right ),
\end{eqnarray}
with $r_{2}$ the \textit{dilaton parameter} given by 
\begin{eqnarray}
r_{2}&=&Q_{_{\text {EMd}}}^{2}e^{2\phi _{0}}/M,
\label{eq:dilParam}
\end{eqnarray}
which involves the mass parameter $M$, the asymptotic value of the dilaton, $\phi _{0}$, as well as the electric charge, $Q_{_{\text {EMd}}}$. In contrast to the RN source, the electric charge in EMd gravity arises from the dilaton-photon coupling. The Schwarzschild space-time is recovered when $Q_{_{\text {EMd}}}=0$.
The EMd solutions are also usually characterized by a \textit{dilaton charge}, $D$, which is given by $D=-Q_{_{\text {EMd}}}^2 e^{2\phi_0}/2M$. This dilaton charge is not a free parameter though, as it depends on $Q_{_{\text {EMd}}}$, $\phi_0$, and $M$.

\section{Markov Chain Monte Carlo method applied to stellar orbits }
\label{sec:MCMC}
The Markov Chain Monte Carlo (MCMC) method is a specific class of Monte Carlo techniques. It yields posterior distributions for the parameters of a given model by sampling from a target distribution using Markov chains. The main difference between pure Monte Carlo methods and MCMC is that, in the latter, the samplings are correlated \citep{speagle2019conceptual, Sharma_2017}.

The MCMC method applied to each space-time model relies on simulations of stellar orbits using the corresponding geodesic equations, and comparing the resulting positions and radial velocities with the available observational data. The comparison between theory and observation proceeds in six steps:
\begin{itemize}
\item[(i)~~] Select a set of values for the free parameters (see Table~\ref{table:Priors}).
\item[(ii)~] Compute the model orbit of the S2 star around Sgr\,A* by integrating the time-like geodesic equations.
\item[(iii)] Project the integrated orbit onto sky coordinates.
\item[(iv)] Interpolate the model orbit positions and velocities to the epochs of the observational data for direct comparison.
\item[(v)~~] Evaluate the likelihood function.
\item[(vi)] Constrain the space-time parameters by accepting or rejecting models according to the $\chi^2$ criterion.
\end{itemize}
Throughout this paper, we use the term \textit{data points} to denote the measured positions and velocities of S2 as observed from Earth, while \textit{points} refers to the positions and velocities generated from the geodesic models.

\subsection{S2 star orbit integration}
The orbit of the S2 star is modelled by a timelike geodesic in the background of a given black hole space-time. For each space-time listed in Section~\ref{sec:spacetimes}, the corresponding second-order ordinary differential equations of motion are derived using the Lagrangian formalism outlined in Appendix~\ref{sec:EulerLagrange}. Given suitable initial conditions (see Section~\ref{sec:initial_conditions}), we numerically integrate these equations to obtain the orbital positions $(t(\tau), r(\tau), \theta(\tau), \phi(\tau))$, parametrized by the proper time of the orbit $\tau$. As initial condition, for simplicity we adopt a turning point of the orbit. Specifically, we use the 2010 apocentre observations and integrate the equations of motion both forward and backward in time, thus covering the full observational time span of the publicly available observational data points.

Since each space-time considered in this work has a Killing vector field $\bm{\xi}_\phi$, the conservation of angular momentum ensures that the orbit is confined to a fixed plane, $\theta=\text{constant}$. In this plane, we adopt Cartesian coordinates defined through the standard transformation $(x,y) = (\bar r \cos \phi, \bar r \sin \phi)$, where we have defined the radial distance $\bar r = \sqrt{h(r)}$ as not for all space-times the radial coordinate coincides with the aerial radius. For each space-time and choice of orbital parameters, we numerically generate stellar orbits represented as sets of points $(x,y)$ in the center-of-mass reference frame, i.e., a coordinate system with its origin at Sgr\,A*.

 \subsection{Orbital precession}
 \label{sec:precession}
 The orbital precession is computed by integrating the equation $d\phi / dr = \dot{\phi} / \dot{r}$. For timelike trajectories in the plane $\theta = \pi/2$, the denominator $\dot{r}$ can be expressed in terms of the metric functions $f(r)$, $g(r)$, and $h(r)$ as follows:
 \begin{eqnarray}
    \dot{r}^{2} = -\frac{1}{g(r)}\left ( \frac{\ell^{2}}{h(r)} + 1\right ) + \frac{1}{g(r)}\frac{f(r_{t})}{f(r)}\left ( \frac{\ell^{2}}{h(r_{t})} + 1\right ),
    \label{eq:rdot}
\end{eqnarray}
where $r_{t}$ is a turning point of the orbit (apocentre or pericentre). Equation \eqref{eq:rdot} follows from the four-velocity normalization condition, $u_{\mu }u^{\mu } = -1$, making use of the conserved quantities of the system: the specific energy $\mathcal{E}$ and the specific angular momentum $\ell$ of the massive test body (see Appendix \ref{sec:EulerLagrange}). In turn, the angular momentum $\ell$ can be expressed as a function of the apocentre, $r_\mathrm{a}$, and pericentre, $r_\mathrm{p}$, through the formula
\begin{eqnarray}
    \ell^{2} = \frac{f(r_\mathrm{a}) - f(r_\mathrm{p})}{\frac{f(r_\mathrm{p})}{h(r_\mathrm{p})} - \frac{f(r_\mathrm{a})}{h(r_\mathrm{a})}}.
    \label{eq:lz}
\end{eqnarray}

Using equations \eqref{eq:rdot} and \eqref{eq:lz}, the orbital precession $\Delta \phi$ for a space-time described by the line element in equation \eqref{eq:generic_line_element} can be computed as
\begin{eqnarray}
    \Delta \phi = 2\int_{r_\mathrm{p}}^{r_\mathrm{a}}\frac{\ell}{h\left ( r \right )\dot{r}}dr,
    \label{eq:orbPre}
\end{eqnarray}
where the factor 2 appears since we are integrating in a full orbit, and the angle traveled when going from pericentre to apocentre is equal to the angle traveled in the opposite direction. In our models, the orbital precession is accounted for by numerically evaluating the integral in Eq. \eqref{eq:orbPre}, rather than employing Post-Newtonian (PN) approximations \citep[see, e.g.,][]{della2022orbital}. We avoid such approximations because the MCMC algorithm explores regions of the parameter space where the precession can deviate significantly from the values predicted at 1PN order. As discussed in Section~\ref{sec:results}, these deviations can have non-negligible consequences.

\subsection{Initial conditions for position and velocity}
\label{sec:initial_conditions}
We use initial conditions corresponding to an osculating Keplerian orbit at the apocentre \citep{della2022orbital}. Since the equations of motion are of second-order, integration requires initial conditions for both position and velocity of the S2 star. The ones related to position are given by
\begin{eqnarray}
t(0)=   t_\mathrm{a},  \qquad 
r(0) = r_\mathrm{a} = a(1+e), \qquad
\phi(0)=\pi,
\label{eq:InitC_A}
\end{eqnarray}
where $t_\mathrm{a}$ is the coordinate time at the orbit's apocentre, $a$ is the semi-major axis, and $e$ is the eccentricity. Similarly, initial conditions for the velocity are given by
\begin{eqnarray}
\dot{t}(0)&=&\sqrt{\frac{h(r_{\mathrm{a}})\dot{\phi}_0^{2} +1}{f(r_{\mathrm{a}})}}, \\
\dot{r}(0)&=&0, \\
\dot{\phi}(0)&=&\dot{\phi}_0=(2 \pi / T)(1-e)^{1 / 2}(1+e)^{-3 / 2},
\label{eq:InitC_B}
\end{eqnarray}
where $T$ is the orbital period, related to the semi-major axis and the central mass $M$ by Kepler's third law, $T^{2}=4\pi^{2}a^{3}/M$. The coordinate time initial rate, $\dot{t}(0)$, comes from the timelike geodesic normalization condition of the four-velocity. 

As in several previous studies of S2 dynamics \citep[e.g.][]{fernandez2023constraining, abuter2020detection}, the Keplerian orbit elements are used here solely as a convenient geometrical parametrization of the osculating Newtonian ellipse at the initial epoch since we are fixing them at apocentre. Kepler's third law is therefore employed only to map these parameters into the coordinate velocity $\dot\phi_0$ at apocentre, and is not assumed to hold as a physical relation in any of the space-times considered.  Once the initial conditions $(r_0,\phi_0,\dot r_0,\dot\phi_0)$ are specified, the subsequent motion is fully determined by the geodesic equations of the metric; hence the choice of this parametrization has no impact on the dynamical evolution. It is also worth mentioning that the point chosen for the initial conditions is the apocentre, where we can expect a weak field limit and thus set the initial conditions with this method \citep[see, e.g.,][]{Li:2025qcv}.

\subsection{Orbit projection to sky coordinates}\label{sec:orbit_projection}
The equations of motion are integrated in the orbital plane, thus the resulting set of coordinates $(x,y)$ must be projected onto the sky plane to compare them directly with observational data. The sky-plane coordinates $(X,Y)$, and a depth coordinate $Z$ are given by
\begin{eqnarray}
X = \mathcal{B}x + \mathcal{G}y, \qquad
Y = \mathcal{A}x + \mathcal{F}y, \qquad
Z = \mathcal{C}x + \mathcal{H}y,
\end{eqnarray}
where $(\mathcal{A},\mathcal{B},\mathcal{C},\mathcal{F},\mathcal{G},\mathcal{H})$ are the Thiele-Innes constants \citep{Green}, defined by
\begin{eqnarray}
\mathcal{A}&=&\tilde{a}(\cos \Omega \cos \omega-\sin \Omega \sin \omega \cos i), \\
\mathcal{B}&=&\tilde{a}(\sin \Omega \cos \omega+\cos \Omega \sin \omega \cos i), \\
\mathcal{C}&=&-\tilde{a}(\sin \omega \sin i), \\
\mathcal{F}&=&-\tilde{a}(\cos \Omega \sin \omega+\sin \Omega \cos \omega \cos i), \\
\mathcal{G}&=&\tilde{a}(-\sin \Omega \sin \omega+\cos \Omega \cos \omega \cos i), \\
\mathcal{H}&=&-\tilde{a}(\cos \omega \sin i),
\end{eqnarray}
where $i$ is the inclination of the orbital plane with respect to the plane of the sky, $\Omega$ is the longitude of the ascending node, and $\omega$ is the perihelion argument of the osculating ellipse associated with the initial condition; $\tilde{a}$ is a scaling factor defined as $\tilde{a}=(3600*180/\pi)/r_d$, with $r_d$ the distance to the Galactic Centre. This scaling factor sets the coordinates $(X,Y,Z)$ in arcseconds, allowing for comparison with the data. Accordingly, the velocity components are
\begin{eqnarray}
V_X=\mathcal{B} v_{\mathrm{x}} +\mathcal{G} v_{\mathrm{y}}, \quad
V_Y=\mathcal{A} v_{\mathrm{x}}+\mathcal{F} v_{\mathrm{y}}, \quad
V_Z=-(\mathcal{C} v_{\mathrm{x}}+\mathcal{H} v_{\mathrm{y}}),
\end{eqnarray}
where the negative sign in the last expression renders $V_Z$ positive when the S2 star approaches Earth.

\subsection{Effects on the apparent orbit}

\subsubsection{R{\o}mer time-delay}
With respect to Earth, the S2 star moves along the $Z$-direction as it orbits Sgr\,A*, causing the light travel time to vary depending on its position. This effect, known as the \textit{R{\o}mer delay} \citep{shea1998ole}, must be included when relating the emission time ($t_{\mathrm{em}}$) to the observation time ($t_\mathrm{obs}$), as expressed by the following relation\footnote{Physical units are restored here for clarity.}
\begin{eqnarray}
t_{\mathrm{obs}}-t_{\mathrm{em}}=\frac{Z\left(t_{\mathrm{em}}\right)}{c}.
\label{eq:RomerD}
\end{eqnarray}
We need to know emission times given observation times of data points in order to compare with integrated orbits, thus we need to solve Eq. \eqref{eq:RomerD} for $t_{\mathrm{em}}$. However, we do not know the function $Z\left(t_{\mathrm{em}}\right)$, so we estimate $t_{\mathrm{em}}$ following the approximation by \cite{S2Redshift},
\begin{eqnarray}\label{eq:t_em}
   t_{\mathrm{em}} \approx t_{\mathrm{obs}}-\frac{Z\left(t_{\mathrm{obs}}\right)}{c} .
\end{eqnarray}

In order to model the observed astrometric positions, $(\bar X, \bar Y)$, we consider a \textit{zero-point offset}, $(x_0, y_0)$, and a \textit{drift}, $(v_{x, 0}, v_{y, 0})$, corresponding, respectively, to the position offset of Sgr\,A*, and the relative velocity of the Galactic Centre, both measured with respect to the reference frame in which observational data were measured. Following the approach of \cite{shaymatov2023parameters}, and using the approximation in Eq.~\eqref{eq:t_em}, the observed astrometric positions are
\begin{eqnarray}
\bar X &=&X\left(t_{\mathrm{em}}\right)+x_0+v_{x, 0}\left(t_{\mathrm{em}}-t_{\mathrm{J} 2000}\right), \label{eq:Xpos} \\
\bar Y &=&Y\left(t_{\mathrm{em}}\right)+y_0+v_{y, 0}\left(t_{\mathrm{em}}-t_{\mathrm{J} 2000}\right),   
\label{eq:Ypos}
\end{eqnarray}
where $t_{\mathrm{J} 2000}$ corresponds to the Julian year 2000.

\subsubsection{Redshift effects}
Among the main contributions to the redshift ($\text{z}$) are the relativistic Doppler effect ($\text{z}_D$) due to the velocity of the S2 star relative to Earth, and the gravitational redshift ($\text{z}_G$) resulting from the star position relative to Sgr\,A*. The cosmological redshift effect is negligible at these scales. The two relevant contributions are given by
\begin{eqnarray}
\text{z}_D &=&\frac{\sqrt{1-V^2\left(t_{\mathrm{em}}\right)/c^2}}{1-\boldsymbol{k} \cdot \boldsymbol{V}\left(t_{\mathrm{em}}\right)},\\
\text{z}_G &=&\frac{1}{\sqrt{f\left(t_{\mathrm{em}}, \boldsymbol{x}_{\mathrm{em}}\right)}},
\label{eq:RedComp}
\end{eqnarray}
where $\boldsymbol{k} \cdot \boldsymbol{V} = V_Z/c$ is the projection of the spatial velocity $\boldsymbol{V}$ of the S2 star along the line of sight. These formulae are evaluated at coordinates corresponding to the signal emission ($t_{\mathrm{em}}, \boldsymbol{x}_{\mathrm{em}}$), so that R{\o}mer delay is directly taken into account. The total redshift is given by
\begin{eqnarray}
    \text{z}=\text{z}_G \text{z}_D-1.
\label{eq:Redshift_addition}
\end{eqnarray}
This allows to quantify the radial velocity of the S2 star, $V_R$, via
\begin{eqnarray}
\text{z}=\frac{\Delta \nu}{\nu}=\frac{\nu_{\mathrm{em}}-\nu_{\mathrm{obs}}}{\nu_{\mathrm{obs}}}=\frac{V_R}{c} .
    \label{eq:Redshift}
\end{eqnarray}
Finally, we consider a constant drift component, $v_{z,0}$, to take into account possible systematic effects on the radial velocity measurement. Similar to astrometric positions in equations \eqref{eq:Xpos}-\eqref{eq:Ypos}, we model the radial velocity as
\begin{eqnarray}
\bar V_R = c  \text{z} +  v_{z,0}.
\label{eq:RavVel}
\end{eqnarray}
Although we will follow the above approach for computing the gravitational redshift, it is worth mentioning that there is another way involving the integration of the $\dot t$ component \citep[see, e.g.,][]{PyGRO}.

\subsection{Model comparison}
Coordinates in the sky plane are compared with the most recent publicly available observational data sets (see Section \ref{sec:datasets}) through the likelihood function
\begin{eqnarray}
    \log \mathcal L =\log \mathcal{L}_{\text {pos}} + \log \mathcal{L}_{V_R} + \log \mathcal{L}_{\text {pre}},
    \label{eq:LogT}
\end{eqnarray}
where $ \log \mathcal{L}_{\text {pos}}$, $\log \mathcal{L}_{V_R}$, and $\log \mathcal{L}_{\text {pre}}$ are logarithmic likelihood functions associated with position, radial velocity, and orbital precession, respectively. These are defined by
\begin{gather}
\log \mathcal{L}_{\text {pos}} = - \frac{1}{2}\sum_i \left[\left( \frac{\bar X^i - X_{\text {obs}}^{i}}{X_{\text {err}}^{i}} \right)^2 + \left( \frac{\bar Y^{i} - Y_{\text {obs}}^{i}}{Y_{\text {err}}^{i}} \right)^2\right], \\
\log \mathcal{L}_{V_R} = -\frac{1}{2}\sum_i\left( \frac{\bar V_R^{i} - V_{R,\text {obs}}^{i}}{V_{R,\text {err}}^{i}} \right)^2, \\
\log \mathcal{L}_{\text {pre}} = -\frac{1}{2} \frac{\left(\Delta \phi - \Delta \phi_{\text {per orbit}}\right)^2}{\sigma_{\Delta \phi, \text {err}}^2}.
\label{eq:likelihood}
\end{gather}
Here, the sum runs over all available observational data points. In $\log \mathcal L_{V_R}$, the term $\bar V_R$ denotes the radial velocity predicted by the model given by Eq. \eqref{eq:RavVel}, and $V_{R,\text {obs}}$, $V_{R,\text {err}}$ correspond to the observed radial velocity and its associated error, respectively. The same conventions apply to $\log \mathcal L_{\text {pos}}$ and $\log \mathcal L_{\text {pre}}$. The orbital precession predicted by the model for a given set of parameters, $\Delta\phi$, is computed by integrating Eq. \eqref{eq:orbPre} from the pericentre $r_{\mathrm{p}} = a(1-e)$ to the apocentre $r_{\mathrm{a}} = a(1+e)$, while the observed value and its error are obtained from Eq. \eqref{eq:PMG}, below.

For the log-likelihood calculation in Eq. \eqref{eq:LogT}, we follow the approach of \cite{yan2024observational}, which, based on the robustness of the precession measurement by the GRAVITY Collaboration (using observational data points also employed in this work), treats the precession term as independent. Incorporating the radial velocity and precession likelihoods provides additional constraints on the models, as opposed to using only the position values as in \cite{bambhaniya2024relativistic}.
To assess the impact of this choice, we perform a robustness test in which all likelihood terms are uniformly downweighted by a factor $\sqrt{2}$ and compare the resulting parameter constraints with the baseline analysis. The central values and $1\sigma$ intervals obtained for the Schwarzschild model under the two likelihood prescriptions are reported in Table~\ref{table:LikelihoodComparison} of Appendix~\ref{sec:likelihood_sensitivity}.

A parameter value is deemed \textit{accepted} if the corresponding likelihood is better than that of the previous iteration, \ie if it is closer to zero. After all realizations, the set of accepted values of a parameter, for any given model, forms a histogram representing its posterior distribution. At the end of the process, we obtain posterior distributions for all the parameters.

\subsection{Orbit interpolation}
Since the equations of motion are integrated numerically, the generated orbits can contain an arbitrary number of points. However, Eq. \eqref{eq:LogT} involves a sum over the exact number of observational data points. Therefore, for each data point, there must be only one corresponding point in the model to calculate the likelihood.

Given the set of observation dates, $\{t^i_\mathrm{obs}\}$, we compute the corresponding set of proper times of the orbit, $\{\tau^i_\mathrm{obs}\}$, by interpolating the numerical solution $t(\tau)$ of the Euler-Lagrange system \eqref{eq:EcsL}, assuming that the coordinate time, $t$, corresponds to the observation dates.
We thus establish a one-to-one correspondence between the model points and the observational data points for each orbit. This correspondence is then used to evaluate Eqs. \eqref{eq:Xpos} - \eqref{eq:Ypos} for astrometric positions and Eq. \eqref{eq:RavVel} for radial velocity. Finally, using Eq. \eqref{eq:LogT}, we compute the likelihood for all orbits. It is worth mentioning that this process of interpolation does not artificially introduce any new data points but rather interpolates points of the model using the dates of the data points as a label for a direct and consistent comparison between model and data through the Likelihood functions.

\subsection{Model parameters and priors}
Each possible orbit of the S2 star is characterized by a minimum number of parameters, which  can be grouped as follows
\begin{equation}
\left\{ \left\{ r_d, M \right\}, \left\{ a, e, t_\mathrm{a}, i, \omega, \Omega \right\}, \left\{ x_0, y_0, v_{x, 0}, v_{y, 0}, v_{z, 0} \right\}, P \right\}.
\label{eq:compSet}
\end{equation}
The first subset of parameters includes the distance from Earth to the Galactic Centre, $r_d$, and the mass of the SMBH Sgr\,A*, $M$. The second subset consists of the Keplerian parameters associated with the initial condition for integrating equations of motion---see Sections \ref{sec:initial_conditions} and \ref{sec:orbit_projection}.
The parameters in the third subset are the offset and drift introduced in equations \eqref{eq:Xpos}, \eqref{eq:Ypos}, and \eqref{eq:RavVel}. The last subset, $P$, corresponds to parameters associated to the particular space-time under consideration; for example, $P = \left\{ Q \right\}$ in the Reissner-Nordstr\"om case.

In the MCMC algorithm we use astrometric and spectroscopic measurements, in addition to the orbital precession measurement by the GRAVITY collaboration.
The parameter space exploration was performed with the Python \textsc{emcee} library \citep{foreman2013emcee}, which implements the MCMC algorithm. For this work, 30 to 32 MCMC chains were used depending on the total number of parameters in each case, using a combination of different types of moves in the parameter space to speedup convergence. Technical details and code validation can be found in Appendix \ref{sec:CodeVal}. All runs were carried out using uniform priors, rather than Gaussian priors, to adopt a questioner approach \citep[see e.g. ][]{della2022orbital}. This choice allows for an agnostic exploration of the parameter space for each space-time as we are not \textit{a priori} expecting small deviations from GR, which is crucial when analysing less studied space-times in this context like the Bardeen black hole or the JNW naked singularity. Specifically, we set our priors so that the posterior distributions feature closed regions corresponding to $3\sigma$ confidence levels, or we adopt values from previous studies when indicated. The priors used in this work are summarized in Table \ref{table:Priors}.

\begin{table}
\centering
\scalebox{1}{
\begin{tabular}{lccccc}
\hline
Parameter          & Lower limit & Upper limit & \multicolumn{1}{c}{Units}          \\ \hline
\addlinespace[3pt]
$r_{d}$      & 7,200 & 9,100 & $\mathrm{pc}$           \\
\addlinespace[3pt]
$M$      & 3.2 & 5.3 & $10^6 \mathrm{M}_{\odot}$                     \\
\addlinespace[3pt]
$a$                & 121 & 134 & $\mathrm{mas}$                     \\
\addlinespace[3pt]
$e$                & 0.872 & 0.898 & -                                  \\
\addlinespace[3pt]
$t_{\mathrm{a}} - 2010$ & 0.330 & 0.465 & $\mathrm{yr}$                      \\
\addlinespace[3pt]
$i$                & 131.4 & 136.0 & ${ }^{\circ}$                      \\
\addlinespace[3pt]
$\omega$           & 62.0 & 68.5 & ${ }^{\circ}$                      \\
\addlinespace[3pt]
$\Omega$           & 223.2 & 229.0 & ${ }^{\circ}$                      \\
\addlinespace[3pt]
$x_0$              & -10 & 10 & $\mathrm{mas}$                     \\
\addlinespace[3pt]
$y_0$              & -10 & 10 & $\mathrm{mas}$                     \\
\addlinespace[3pt]
$v_{x, 0}$         & -1 & 1 & $\mathrm{mas} \, \mathrm{yr}^{-1}$ \\
\addlinespace[3pt]
$v_{y, 0}$         & -1 & 1 & $\mathrm{mas} \, \mathrm{yr}^{-1}$ \\
\addlinespace[3pt]
$v_{z,0}$ & -60 & 60 & $\mathrm{km} \, \mathrm{s}^{-1}$   \\ 
\addlinespace[3pt]
\hline
\addlinespace[3pt]
$Q$ & -45 & 45 & $ 10^{6}\mathrm{M}_{\odot}$ \\
\addlinespace[3pt]
$q_{_\text{JNW}}$ & -600 & 600 & $ 10^{6}\mathrm{M}_{\odot}$ \\
\addlinespace[3pt]
$g$ & -260 & 260 & $ 10^{6}\mathrm{M}_{\odot}$ \\
\addlinespace[3pt]
$q_{\text{H}}$ & -6.9 $\times 10^{-3}$ & 8.4 $\times 10^{-3}$ & $ 10^{6}\mathrm{M}_{\odot}$ \\
\addlinespace[3pt]
$\delta$ & -0.99 & 4 & - \\
\addlinespace[3pt]
$\lambda$ & $10^{4}$ & $5\times 10^{6}$ & $ 10^{6}\mathrm{M}_{\odot}$ \\
\addlinespace[3pt]
$\phi_0$ & 0 & 5 & - \\
\addlinespace[3pt]
$Q_{_{\text {EMd}}}$ & -9 & 9 & $ 10^{6}\mathrm{M}_{\odot}$ \\
\addlinespace[3pt]
\hline
\end{tabular}
}
\caption{The upper part of the table contains a list of uniform priors for the 13 parameters common to all the models considered in this work. The lower part is a list of uniform priors for parameters specific to each model, in geometrized units.}
\label{table:Priors}
\end{table}

\begin{table*}
\resizebox{\textwidth}{!}{%
\begin{tabular}{lcccccccc}
\hline
\multicolumn{1}{c}{Parameter} & \multicolumn{1}{c}{Schwarzschild} & \multicolumn{1}{c}{RN} & \multicolumn{1}{c}{JNW} & \multicolumn{1}{c}{Bardeen} & \multicolumn{1}{c}{Horndeski} & \multicolumn{1}{c}{Yukawa-like} & \multicolumn{1}{c}{EMd}  \\
\hline

\addlinespace[5pt] 
    
$r_{d}$ (pc)                  
& $8,115^{+220}_{-217}$          
& $8,132^{+217}_{-218}$                             
& $8,128^{+220}_{-217}$
& $8,132^{+218}_{-217}$ 
& $8,117^{+218}_{-217}$ 
& $8,119^{+222}_{-218}$ 
& $8,129^{+222}_{-219}$\\
\addlinespace[5pt]

$M$ ($10^6 \mathrm{M}_{\odot}$)                  
& $4.236^{+0.231}_{-0.217}$     
& $4.255^{+0.230}_{-0.220}$                             
& $4.250^{+0.232}_{-0.219}$
& $4.254^{+0.231}_{-0.220}$ 
& $4.239^{+0.230}_{-0.219}$ 
& $4.242^{+0.235}_{-0.220}$ 
& $4.251^{+0.233}_{-0.220}$\\

\addlinespace[5pt]

$a$ (mas)                            
& $127.20^{+1.35}_{-1.27}$      
& $127.13^{+1.35}_{-1.25}$                             
& $127.23^{+1.34}_{-1.27}$
& $127.12^{+1.34}_{-1.26}$ 
& $127.16^{+1.35}_{-1.26}$ 
& $127.17^{+1.35}_{-1.28}$ 
& $127.14^{+1.36}_{-1.28}$\\

\addlinespace[5pt]

$e$                            
& $0.88480^{+0.00274}_{-0.00270}$
& $0.88471^{+0.00274}_{-0.00267}$                             
& $0.88456^{+0.00273}_{-0.00271}$
& $0.88469^{+0.00272}_{-0.00266}$ 
& $0.88472^{+0.00273}_{-0.00266}$ 
& $0.88479^{+0.00274}_{-0.00271}$ 
& $0.88473^{+0.00275}_{-0.00271}$\\

\addlinespace[5pt]

$t_{\mathrm{a}} - 2010$ (yr)       
& $0.3972^{+0.0137}_{-0.0135}$                             
& $0.3974^{+0.0138}_{-0.0135}$                             
& $0.3973^{+0.0138}_{-0.0136}$
& $0.3974^{+0.0138}_{-0.0136}$ 
& $0.3972^{+0.0137}_{-0.0136}$ 
& $0.3968^{+0.0138}_{-0.0135}$ 
& $0.3974^{+0.0138}_{-0.0136}$\\

\addlinespace[5pt]

$i$ ($^{\circ}$)                            
& $133.80^{+0.51}_{-0.51}$                                  
& $133.83^{+0.50}_{-0.51}$                                       
& $133.82^{+0.51}_{-0.51}$
& $133.83^{+0.50}_{-0.51}$ 
& $133.81^{+0.50}_{-0.51}$ 
& $133.81^{+0.51}_{-0.52}$ 
& $133.83^{+0.51}_{-0.52}$\\

\addlinespace[5pt]

$\omega$ ($^{\circ}$)                       
& $65.26^{+0.70}_{-0.70}$                                   
& $65.23^{+0.71}_{-0.69}$                                       
& $65.23^{+0.70}_{-0.70}$
& $65.22^{+0.70}_{-0.69}$ 
& $65.25^{+0.70}_{-0.69}$ 
& $65.25^{+0.70}_{-0.71}$ 
& $65.23^{+0.71}_{-0.71}$\\

\addlinespace[5pt]

$\Omega$ ($^{\circ}$)                       
& $226.06^{+0.71}_{-0.70}$                                  
& $226.03^{+0.71}_{-0.70}$                                       
& $226.04^{+0.71}_{-0.70}$
& $226.02^{+0.71}_{-0.70}$ 
& $226.05^{+0.71}_{-0.70}$ 
& $226.06^{+0.71}_{-0.71}$ 
& $226.03^{+0.71}_{-0.71}$\\

\addlinespace[5pt]

$x_0$ (mas)                          
& $0.14^{+0.44}_{-0.44}$                                  
& $0.15^{+0.44}_{-0.44}$                                       
& $0.15^{+0.44}_{-0.44}$
& $0.15^{+0.44}_{-0.44}$ 
& $0.09^{+0.44}_{-0.44}$ 
& $0.11^{+0.44}_{-0.45}$ 
& $0.16^{+0.44}_{-0.44}$\\

\addlinespace[5pt]

$y_0$ (mas)                          
& $-2.10^{+0.68}_{-0.69}$                                  
& $-2.06^{+0.67}_{-0.69}$                                       
& $-2.07^{+0.69}_{-0.69}$
& $-2.07^{+0.68}_{-0.69}$ 
& $-2.09^{+0.68}_{-0.69}$ 
& $-2.11^{+0.69}_{-0.69}$ 
& $-2.07^{+0.69}_{-0.70}$\\

\addlinespace[5pt]

$v_{x, 0}$ (mas yr$^{-1}$)                     
& $0.124^{+0.046}_{-0.046}$                                  
& $0.122^{+0.046}_{-0.046}$                                       
& $0.122^{+0.046}_{-0.046}$
& $0.122^{+0.046}_{-0.046}$ 
& $0.128^{+0.046}_{-0.046}$ 
& $0.126^{+0.046}_{-0.046}$ 
& $0.122^{+0.046}_{-0.046}$\\

\addlinespace[5pt]

$v_{y, 0}$ (mas yr$^{-1}$)                     
& $0.012^{+0.072}_{-0.072}$                                  
& $0.008^{+0.072}_{-0.072}$                                       
& $0.008^{+0.072}_{-0.072}$
& $0.008^{+0.072}_{-0.072}$ 
& $0.014^{+0.073}_{-0.072}$ 
& $0.013^{+0.073}_{-0.073}$ 
& $0.008^{+0.072}_{-0.072}$\\

\addlinespace[5pt]

$v_{z, 0}$ (km s$^{-1}$)                     
& $24.4^{+8.0}_{-7.9}$ 
& $24.6^{+8.0}_{-7.9}$ 
& $24.5^{+7.9}_{-7.9}$
& $24.7^{+7.9}_{-7.9}$ 
& $24.4^{+7.9}_{-7.9}$ 
& $24.6^{+8.0}_{-7.9}$ 
& $24.6^{+8.0}_{-8.0}$\\

\addlinespace[5pt]
\hline
\addlinespace[5pt]
$Q$ ($10^6 \mathrm{M}_{\odot}$) 
& - 
& $0.01^{+10.01}_{-10.03}$ 
& - 
& - 
& - 
& - 
& -  \\

\addlinespace[5pt]

$q_{_\text{JNW}}$ ($10^6 \mathrm{M}_{\odot}$) 
& - 
& - 
& $0.40^{+134.59}_{-135.50}$
& - 
& - 
& - 
& -  \\

\addlinespace[5pt]

$g$ ($10^6 \mathrm{M}_{\odot}$) 
& - 
& - 
& - 
& $0.12^{+68.99}_{-68.66}$ 
& - 
& - 
& -  \\

\addlinespace[5pt]

$q_\text{H}$ ($10^6 \mathrm{M}_{\odot}$) 
& - 
& - 
& - 
& - 
& $0.0009^{+0.0013}_{-0.0013}$ 
& - 
& -  \\

\addlinespace[5pt]

$\delta$ 
& - 
& - 
& - 
& - 
& - 
& $\gtrsim -0.04$ 
& -  \\

\addlinespace[5pt]

$\lambda$ ($10^6 \mathrm{M}_{\odot}$) 
& - 
& - 
& - 
& - 
& - 
& $2.67^{+1.24}_{-0.62} \times 10^{6}$  
& -  \\

\addlinespace[5pt]

$\phi_{0}$ 
& - 
& - 
& - 
& - 
& - 
& - 
& $\lesssim 1.82$\\
\addlinespace[5pt]

$Q_{_{\text {EMd}}}$ ($10^6 \mathrm{M}_{\odot}$) 
& - 
& - 
& - 
& - 
& - 
& - 
& $0.00^{+1.33}_{-1.31}$\\

\addlinespace[5pt]
\hline
\end{tabular}%
}
\caption{Mean value and 1$\sigma$ bound from the MCMC method. The upper part of the table shows the constraints to the 13 parameters common to all models. The lower part shows the constraints to the distinctive parameters of each model, in geometrized units.}
\label{table:BestFit}
\end{table*}

\subsection{Observational data set}
\label{sec:datasets}
In order to constrain our theoretical models with observations, we made use of the most complete, publicly available data sets for the S2 star orbiting Sgr\,A*. The observations used in our analysis provide three independent probes of the gravitational field:
\begin{itemize}
    \item[(i)~~] Astrometric positions of the S2 star: 145 data points from the years $\sim$ 1992 to $\sim$ 2016, where data before 2002 were measured by the System for High Angular Resolution Pictures (SHARP) \citep{hofmann1993high}, and the rest with the Nasmyth Adaptive Optics System (NACO) \citep{rousset1998design,lenzen1998conica}.
    \item[(ii)~]  Radial velocities of the S2 star: 44 data points from the years $\sim$ 2000 to $\sim$ 2016. In this case, data before 2003 were collected with the Near-Infrared Camera 2 (NIRC2) \citep{ghez2003first}, and the rest were collected with the adaptive-optics-assisted integral field spectrograph Single Faint Object Near-Infrared Investigation (SINFONI) \citep{eisenhauer2003sinfoni, bonnet2004first}.
    \item[(iii)]  Orbital precession of the S2 star: Pericentre advance measurement by the GRAVITY Collaboration \citep{abuter2020detection}. In arcminutes, their measurement amounts to
    \begin{eqnarray}
        \Delta \phi_{\text {per orbit }} = 12.1' \times (1.10 \pm 0.19).
        \label{eq:PMG}
    \end{eqnarray}
\end{itemize}

The astrometric and spectroscopic data used for orbital fits were retrieved from Table~5 of \cite{gillessen2017update}. Since the orbital period of the S2 star is $\sim16$ years, these data cover more than one complete orbit, allowing a direct comparison between the long-term orbital dynamics predicted by theory and observations.

\section{Results} 
\label{sec:results}

\subsection{Posterior distributions} 
In Figures \ref{fig:SchCorner} - \ref{fig:EMD_Corner}, we present corner plots for the seven representative black hole space-times explored in this work. Each panel shows the posterior probability distributions of the parameters describing the S2 star orbit, with colours indicating the 68\% (1$\sigma$), 95\% (2$\sigma$), and 99.7\% (3$\sigma$) confidence levels.
The corresponding best-fit values are summarized in Table \ref{table:BestFit}, where we report the mean values together with their 68\% (1$\sigma$) confidence intervals obtained from the MCMC algorithm.

Figure \ref{fig:SchCorner} shows the posterior distributions obtained for the Schwarzschild space-time, which we adopt as the fiducial model. In other space-times considered, only slight deviations from the corresponding parameter values of this reference model were found. For instance, the inferred mean mass of Sgr\,A* in our MCMC analysis within the seven space-times considered (see Table \ref{table:BestFit}) lies in the range $(4.236 - 4.255) \times 10^6 M_{\odot}$, while its distance to Earth is between $(8,115 - 8,132)$ pc. This corresponds to deviations of at most $0.22 \%$ and $0.10 \%$, respectively. 

Overall, the parameters in the subset $P$ of \eqref{eq:compSet}---corresponding to the specific quantities characterizing each space-time---exhibit only weak or negligible correlations with the other parameter sets across the space-times considered (see, for example, the last row of figures in Appendix \ref{sec:Corn}). 

{\it Reissner-Nordstr\"om black hole.} 
The posterior distributions for the RN space-time are shown in Fig. \ref{fig:RN_Corner}, where the electric charge associated with Sgr\,A* exhibits a symmetric distribution centered at zero, with $Q$ bounded within 1$\sigma$ between
\begin{eqnarray}
-10.02 < Q(10^{6} \mathrm{M}_{\odot}) < 10.02.
\label{eq:ConstrRN}
\end{eqnarray}
This constraint is comparable to the upper limit obtained by \cite{2024MNRAS.530.3038M}, where the authors estimated $\left | Q/M \right | < 2.32$ (with $M=4.29\times10^{6}M_\odot$) by comparing EHT observations with the zero-gravity radius, defined as the radius where a test particle remains at rest and whose existence is a general property of the RN metric in its naked-singularity regime. In our case, Eq. \eqref{eq:ConstrRN} yields a similar bound, $\left | Q/M \right | < 2.34$. Less conservative constraints have been reported using alternative methods. For instance, \cite{2018MNRAS.480.4408Z} proposed an observational test based on the suppression of bremsstrahlung surface brightness, obtaining an upper limit on the charge of Sgr\,A* of $\lesssim 3\times10^{8}$ C. In contrast, Eq. \eqref{eq:ConstrRN} expressed in the same units gives $\left | Q \right | < 4.84\times10^{26}$ C, which is consistent with the estimate $\left | Q \right | \lesssim 3.6\times10^{27}$ C derived by \cite{2012GReGr..44.1753I} from the orbital motion of the S2 star.

{\it Janis-Newman-Winicour and Bardeen black holes.} The posterior distributions for the parameters of the JNW and Bardeen space-times are shown in Figs.~\ref{fig:JNW_Corner} and \ref{fig:Bardeen_Corner}, respectively. At the $1\sigma$ level, we obtain the following bounds for the scalar charge ($q_{_\text{JNW}}$) of the massless scalar field, and for the magnetic monopole charge ($g$),
\begin{eqnarray}
-135.10 < q_{_\text{JNW}}(10^{6} \mathrm{M}_{\odot}) < 134.99,\\
-68.54 < g(10^{6} \mathrm{M}_{\odot}) < 69.11.     \label{eq:ConstBardeen}
\end{eqnarray}
A recent study by \cite{bambhaniya2024relativistic} constrained the scalar charge of the JNW space-time to $\log q_{\text{JNW}} (M{\odot}) = -7.46^{+0.58}_{-0.57}$ at the $2\sigma$ level. However, their analysis only considered the astrometric positions of the S2 star, and excluded a priori the possibility of $q{_\text{JNW}} = 0$. In contrast, the magnetic monopole charge $g$ of the rotating Bardeen space-time was constrained by \cite{2023MNRAS.524.3683A} using EHT observables of the shadow of Sgr\,A*, yielding $g = 0.40^{+0.08}_{-0.11}M$ and a spin parameter $a = 0.65^{+0.10}_{-0.10}M$, with $M = 4.0 \times 10^{6} M_\odot$. Although Eq.~\eqref{eq:ConstBardeen} provides a less restrictive bound, it represents the first constraint on the magnetic charge of the non-rotating Bardeen space-time obtained from observational data of the S2 star.

{\it Hairy black hole from Horndeski gravity.} For the scalar charge $q_\text{H}$ of the Horndeski space-time, the posterior distribution shown in Fig.~\ref{fig:Horndeski_Corner} is approximately Gaussian, centered on a positive value. However, it remains consistent with zero at the $1\sigma$ level, with the corresponding bounds in this region given by
\begin{eqnarray}
 -0.0004 < q_{\text{H}}(10^{6} \mathrm{M}_{\odot}) < 0.0022.  
 \label{eq:HorndeskiCons}
\end{eqnarray}
This range is in agreement with the one obtained by \cite{2023EPJC...83..311L} using only the S2 star orbital precession inferred by the Gravity Collaboration, given in Eq. \eqref{eq:PMG}, and the best-fit orbit parameters from \cite{abuter2020detection}. Their range is given by $\tilde{q} \in [-0.75\times 10^{-4}, 2.395\times 10^{-4}]$, where $\tilde{q}=q_{H}/2M$, and $M=4.261\times 10^{6}\text{M}_{\odot}$. Similarly, Eq. \eqref{eq:HorndeskiCons}  can be rewritten as $\tilde{q} \in [-0.469\times 10^{-4}, 2.582\times 10^{-4}]$.\\

{\it Yukawa-like black hole from $f(\mathcal R)$ gravity.} In this case, the posterior distributions of the parameters $\lambda$ and $\delta$ are shown in Fig.~\ref{fig:Yukawa_corner}. As in \cite{de2021f}, we adopt flat priors; however, we do not use the approximation $\Psi \sim \Phi$. We avoid this simplification because, in our parameter space, the discrepancy between $\Psi$ and $\Phi$ exceeds the $\sim20\%$ deviation reported in \cite{Yukawa2}. Within our model, the length scale $\lambda$ can be constrained at the $1\sigma$ level, while for the Yukawa strength $\delta$ we can only establish a lower bound, namely:
\begin{eqnarray}
\delta \gtrsim  -0.04 ,     
\end{eqnarray}
\begin{eqnarray}
2.05 \times 10^{6} < \lambda (10^{6} \mathrm{M}_{\odot}) < 3.91 \times 10^{6} .
\label{eq:LamdCons}
\end{eqnarray}
In contrast, \cite{de2021f} obtained the constraints $\delta = -0.01^{+0.61}_{-0.14}$ and a lower bound for $\lambda$ at the $1\sigma$ level, $\lambda \gtrsim 6,300$ AU, by incorporating the orbital precession measurement into the likelihood calculation. For comparison, our bounds in Eq.\eqref{eq:LamdCons} correspond to $20,200 < \lambda (\text{AU}) < 38,600$. We stress that our model does not assume $\Psi \sim \Phi$, and that we compute the orbital precession by directly integrating Eq.\eqref{eq:orbPre}, rather than using the approximation adopted by \cite{de2021f}, which leads to larger uncertainties in the shared parameter space of $\delta$ and $\lambda$. Nevertheless, if we do adopt the $\Psi \sim \Phi$ approximation and follow the same precession treatment as in \cite{de2021f}, we obtain bounds consistent with their results.

\begin{figure}
    \centering
    \includegraphics[scale=0.37]{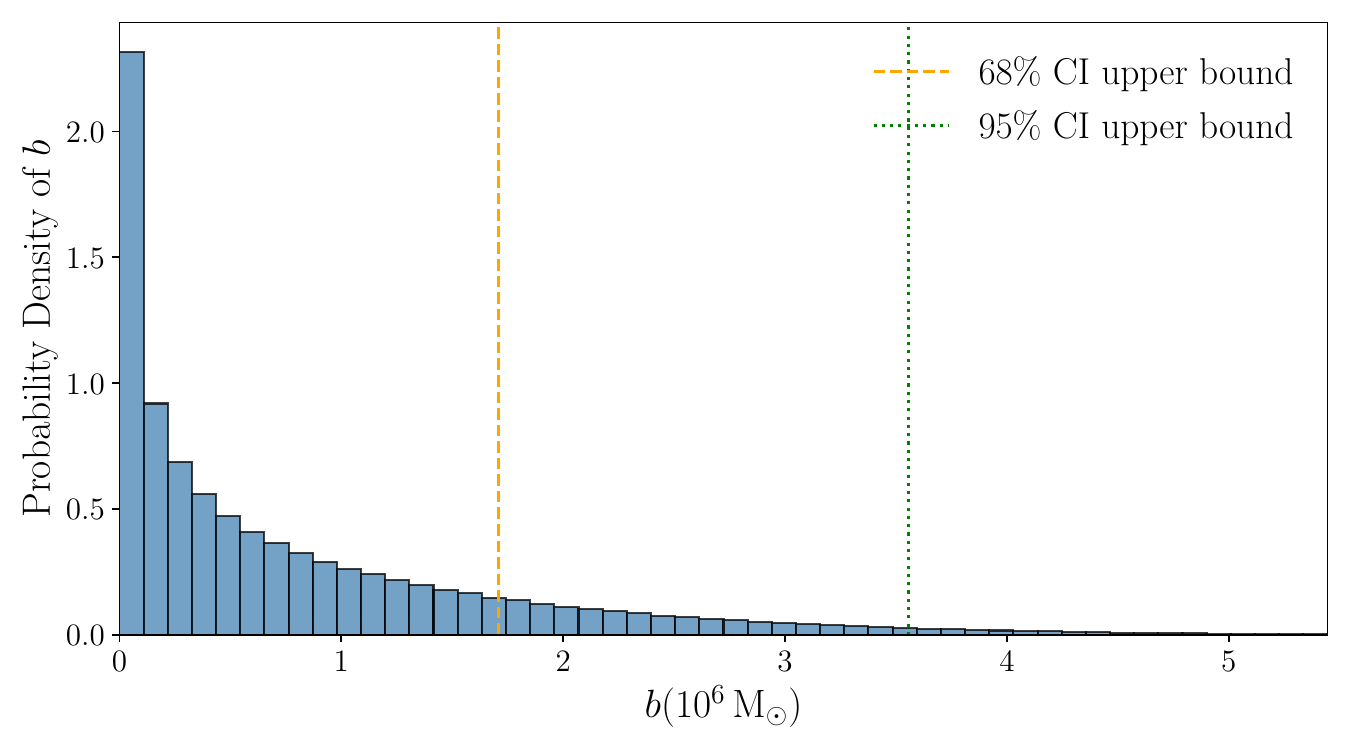}
    \caption{Probability density distribution of the dilaton parameter $b$ of the EMd model. The vertical lines denote the upper limit of $b$ under 68\% and 95\% confidence intervals (CI), which correspond to $1.708\times 10^{6}\ \mathrm{M}_{\odot}$ and $3.555\times 10^{6} \mathrm{M}_{\odot}$, respectively. }
    \label{fig:Marg_b}
\end{figure}

{\it Einstein-Maxwell-dilaton black hole.} Finally, for the electrically charged black hole solution in EMd gravity, the distributions are presented in Fig.~\ref{fig:EMD_Corner}. For the asymptotic value of the dilatonic field, $\phi_0$, the distribution peaks near zero, allowing us to establish an upper bound. For the electric charge, $Q_{_{\text {EMd}}}$, we obtain a $1\sigma$ constriction---one order of magnitude smaller than in the RN model---also centered around zero. These bounds are
\begin{eqnarray}
\phi_0 \lesssim 1.82,
\end{eqnarray}
\begin{eqnarray}
-1.31 < Q_{_{\text {EMd}}} (10^{6} \mathrm{M}_{\odot}) < 1.33.
\end{eqnarray}
Our results can be compared with those obtained by \cite{fernandez2023constraining}, where the value of a rescaled dilaton parameter was constrained, which is defined as $b\equiv r_2/2$, with $r_2$ given in Eq. \eqref{eq:dilParam}. For this comparison, we plot the marginalized posterior distribution of the dilaton parameter $b$ in Fig. \ref{fig:Marg_b}, and calculate the upper limits of the 68\% and 95\% confidence intervals. \cite{fernandez2023constraining} constrain this parameter to $b\lesssim0.058$ AU at the 95\% confidence level---equivalently, $b\lesssim1.4M$ with $M=4.2\times 10^{6} \mathrm{M}_{\odot}$ in the geometrized units used in their study. In our analysis, the upper limit is reduced to $b\lesssim0.035$ AU ($b\lesssim0.85M$), which lies within the range of allowed black hole solutions in this theory, $0\leq b\leq M$. 

\subsection{Model comparison}
\begin{table}
    \centering
    \begin{tabular}{lcccc}
        \hline
        \addlinespace[1pt]
        Model & \multicolumn{2}{c}{Standard Calculation} & \multicolumn{2}{c}{Downweighted Outliers} \\
        \cmidrule(lr){2-3} \cmidrule(lr){4-5}
        & $\chi^{2}$ & $\chi_{\nu}^{2}$ & $\chi^{2}$ & $\chi_{\nu}^{2}$ \\
        \hline
        \addlinespace[1pt]
        Schwarzschild &  289.718 & 1.637 & 277.958 & 1.570 \\
        RN     & 289.718 & 1.646 & 277.958 & 1.579 \\
        JNW    & 289.718 & 1.646 & 277.958 & 1.579 \\
        Bardeen   & 289.718 & 1.646 & 277.958 & 1.579 \\
        Horndeski   & 289.286 & 1.644 & 277.552 & 1.577 \\
        Yukawa-like  & 288.918 & 1.651 & 277.196 & 1.584 \\
        EMd    & 289.718 & 1.656 & 277.958 & 1.588  \\
        \hline
    \end{tabular}
    \caption{Minimum $\chi^2$ and reduced $\chi^2_\nu$ values for each space-time model. The first two columns show results from the standard calculation, while the last two columns correspond to values obtained using a scheme that downweights outliers.}
    \label{table:chiSqr}
\end{table}

We have analyzed our space-time models with the Python library \textsc{dynesty} \citep{speagle2020dynesty, koposov2023joshspeagle}, which implements a Nested Sampling algorithm to evaluate Bayesian evidence. This method also provides an estimation of the maximum likelihood, from which we calculate the $\chi^2$ and reduced $\chi_\nu^2$, as reported in Table \ref{table:chiSqr}. In addition, we minimized $\chi^2$ with an outlier down-weighting scheme following \cite{abuter2020detection}, replacing the standard penalty function $p(r) = r^2$ with $p(r, s) = r^2 s^2 / (r^2 + s^2)$, using $s = 10$ to introduce a smooth cut-off around $10\sigma$ and thereby moderate the influence of each data point. The results, summarized in Table \ref{table:chiSqr}, show that in both approaches the Schwarzschild space-time yields the smallest reduced $\chi^2_\nu$. 

\begin{figure*}
  \begin{center}
   \includegraphics[width=0.73\textwidth]{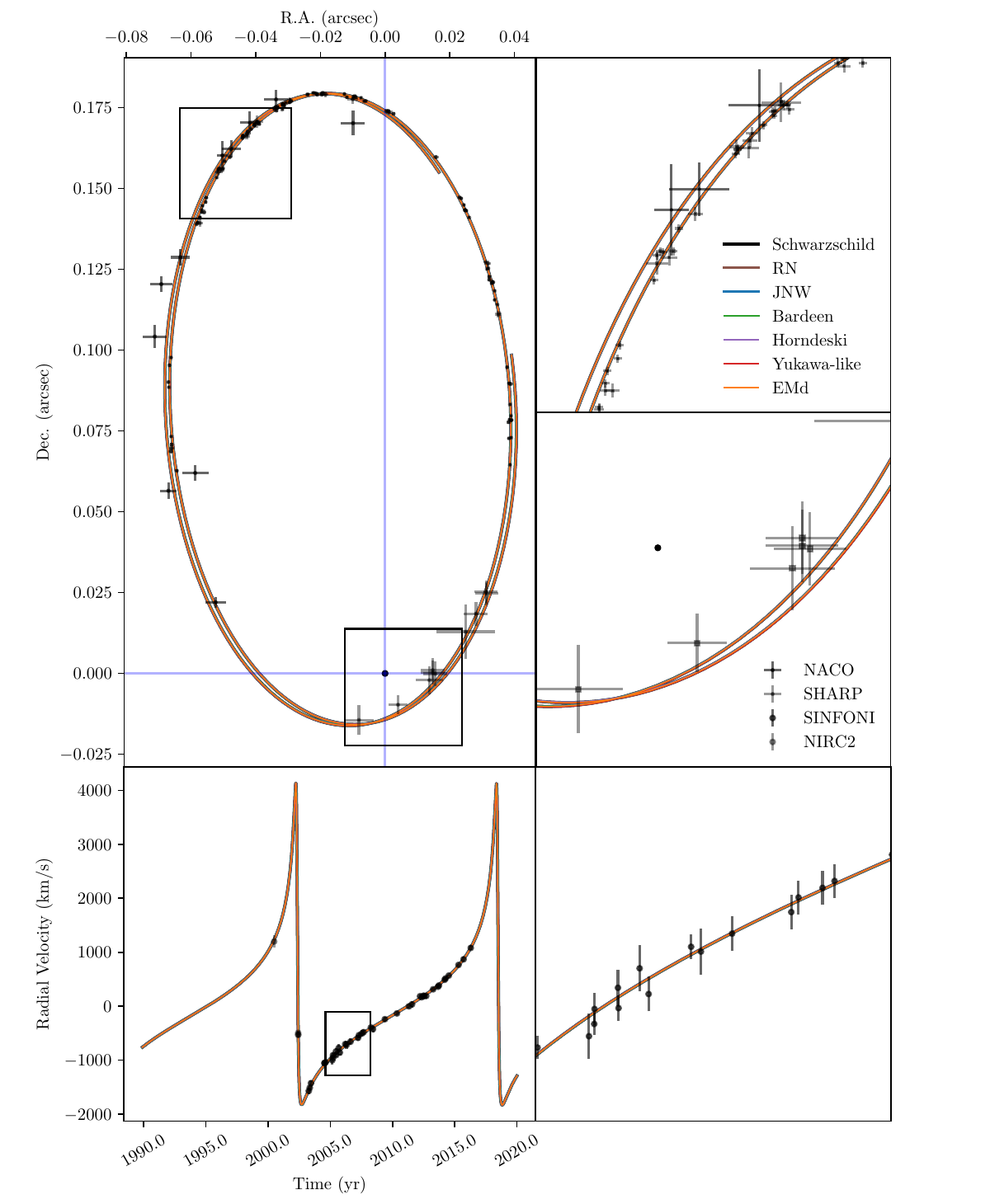}
   \caption{Trajectory in the sky plane and radial velocity of S2 in the different space-time models under consideration. The left part of the figure shows the best-fit trajectory and radial velocity of S2, with the corresponding observational data points, i.e., 145 astrometric positions of the S2 star from the years $\sim$ 1992 to $\sim$ 2016, measured with the SHARP and NACO instruments, and 44 radial velocities from the years $\sim$ 2000 to $\sim$ 2016, measured with the NIRC2 and SINFONI instruments. The right side of the figure shows enlarged views of the boxed regions highlighted on the left. All best-fit models predict a noticeable precession in the orbit of S2. However, no substantial differences between models are discernible.}
   \label{fig:OrbitModels}
  \end{center}
 \end{figure*}

To illustrate how well each model fits the observational data, we plotted the sky plane trajectory and radial velocity using the best-fit parameters associated with the standard $\chi^2$ calculation in Table \ref{table:chiSqr}. The result can be seen in Fig. \ref{fig:OrbitModels}, where the left side of the figure shows the radial velocity in the lower corner and the orbit in the plane of the sky in the upper corner, while the right side of the figure shows the zoomed-in areas marked with boxes over the plots on the left. As shown in the figure, the curves for the different models exhibit no significant differences.

\begin{table*}
    \centering
    \begin{tabular}{lcccccc}
        \hline
        \addlinespace[1pt]
        Model & Parameters  
        & \multicolumn{2}{c}{Standard Calculation} 
        & \multicolumn{2}{c}{Downweighted Outliers} 
        & Strength of evidence \\
        \cmidrule(lr){3-4} \cmidrule(lr){5-6}
        &  & ln\,(Evidence) & ln\,(BF) & ln\,(Evidence) & ln\,(BF) &  \\
        \hline
        \addlinespace[1pt]
        Schwarzschild & 13 & $-182.649 \pm 0.088$ & -  & $-176.360 \pm 0.088$  & -   & Reference Model \\
        Bardeen   & 14    & $-183.400 \pm 0.089$ & $0.751\pm 0.125$ & $-177.082 \pm 0.088$  & $0.722 \pm 0.124$  & Undecided  \\ 
        Yukawa-like  & 15 & $-183.511 \pm 0.089$ & $0.862\pm 0.125$ & $-177.003 \pm 0.088$  & $0.643 \pm 0.124$  & Undecided \\
        JNW    & 14       & $-183.673 \pm 0.089$ & $1.024\pm 0.125$ & $-177.339 \pm 0.089$  & $0.979 \pm 0.125$  &  Weakly disfavoured \\
        Horndeski  & 14  & $ -183.722 \pm 0.089$ & $1.073\pm 0.125$ & $ -177.479\pm 0.089$  & $1.119 \pm 0.125$  & Weakly disfavoured \\
        RN     & 14       & $-183.774 \pm 0.089$ & $1.125 \pm 0.125$ & $-177.380 \pm 0.089$  & $1.020 \pm 0.125$  & Weakly disfavoured  \\
        EMd    & 15       & $-184.905 \pm 0.091$ & $2.256\pm 0.127$ & $-178.634 \pm 0.091$  & $2.274 \pm 0.127$  & Weakly disfavoured  \\
        \hline
    \end{tabular}
    \caption{Bayesian comparison of models, using log-evidence and log-Bayes factors (ln BF) relative to the Schwarzschild model. The first two columns correspond to the standard nested sampling results, while the next two columns show values obtained using a scheme that downweights outliers. The last column categorizes the strength of evidence, based on the central value of the standard calculation, according to Jeffreys' scale.}
    \label{table:log_evidence_comparison}
\end{table*}

It is worth noting that, in the standard calculation of $\chi^2$ and $\chi^2_\nu$, our overall fit for most of the space-times considered yields a reduced $\chi_\nu^2$ comparable to that obtained by \cite{abuter2020detection} ($\chi_\nu^2 \approx 1.65$), where a post-Newtonian limit of the equations of motion \citep{will2008testing} together with a parametrization of the Schwarzschild precession were used. In contrast, applying the outlier down-weighting scheme in our analysis does not decrease the reduced $\chi^2$ values to $\chi_\nu^2 \approx 1.50$ as in their case. We attribute such a difference to the distinct data sets employed.

In addition, we performed a Bayesian comparison of models using the same two previous approaches. Results are reported in Table \ref{table:log_evidence_comparison}. The Bayes factor (BF) was computed from the Bayesian evidence of each model, using the uniform priors listed in Table \ref{table:Priors}.
In this context, Bayesian evidence, or simply \textit{evidence}, provides a measure of the data preference for a particular model \citep{trotta2008bayes}. In the literature, it is also referred to as \textit{marginal likelihood}, defined as the average of the likelihood over the prior. It provides a direct measure of the relative support that the data give to each model.

In our analysis, both the standard calculation of $\chi^2$ and $\chi^2_\nu$, and the outlier down-weighting scheme, yield similar results, where the highest value of log-evidence obtained corresponds, in both cases, to the Schwarzschild space-time, which is therefore used as a reference to calculate the Bayes factor of all other models. For the interpretation of the Bayes factor, we used Jeffreys' scale \citep{jeffreys1998theory} for the strength of evidence. In particular, we rely on the version in terms of natural logarithm shown in \cite{trotta2008bayes}.

It is important to emphasize that the Bayes factors reported here are inherently prior-dependent because the evidence integrates the likelihood over the full prior volume. Since we employ broad, uniform priors (Table \ref{table:Priors}), the resulting Bayes factors naturally reflect this choice, and values with $ \ln(\mathrm{BF}) < 2.5$ should not be interpreted as statistically meaningful preferences \citep{trotta2008bayes}. Although exploring narrower or Gaussian priors could reduce prior-volume effects, such an analysis is beyond the scope of the present work; instead, we explicitly acknowledge this limitation and interpret the Bayes factors with appropriate caution.

Under our priors, and the two taken approaches, the comparison with respect to the benchmark (Schwarzschild) model yields the following results. For the Bardeen and Yukawa-like black holes, the outcome is not conclusive, so we have an {\it undecided result}, indicating that the current data do not provide sufficient discriminatory power to distinguish these models from the benchmark. Similarly, the RN, JNW, EMd, and Horndeski black holes do not show statistically significant evidence of being disfavoured relative to the Schwarzschild space-time; although some models present positive $\ln(\mathrm{BF})$ central values, these differences are small (mostly $<2$) and prior-sensitive. The EMd model shows a central $\ln(\mathrm{BF})$ nearer the conventional threshold, but its strength of evidence falls into the same category as the previous cases. Overall, the present S2 data are statistically consistent with the Schwarzschild geometry and do not provide decisive discrimination among the alternative space-times considered here.

\section{Conclusions} 
\label{sec:conclusion}
We modelled the orbit of the S2 star around Sgr\,A*, by integrating geodesic equations, within seven representative, non-rotating black hole space-times from GR and beyond. Novelties with respect to previous works are the following. $(i)$ We have used a consistent framework across all space-times; $(ii)$ we have taken into account the precession of the S2 star in the analysis for the JNW solution; $(iii)$ for the Yukawa-like solution, we do not restrict ourselves to the approximation $\Psi \sim \Phi$; and $(iv)$ we have analysed the Bardeen regular black hole solution for the first time using observational information of the S2 star orbit.

We then carried out MCMC simulations to compare theoretical predictions against the publicly available observational data of the S2 star. Posterior probability distributions were obtained for the parameters of the seven representative models, allowing us to place constraints at 1$\sigma$, 2$\sigma$, and 3$\sigma$ confidence levels.

Our results show that, in the Bayesian comparison with respect to the Schwarzschild space-time, the observational data indicate no statistically significant preference among any of the models considered. For example, although the EMd space-time yields the largest Bayes factor in our analysis, its value remains within the range of weak evidence.
Overall, within the prior space adopted, features such as electric charge, scalar charge, or parameters from theories beyond GR, cannot yet be strictly excluded with the current precision of the S2 star orbital data.

Among the alternative non-rotating black hole solutions, we find that the Horndeski scalar charge $q_{\text{H}}$ shows a slight preference for positive values, though it remains consistent with zero within the 1$\sigma$ confidence interval. Notably, incorporating the orbital precession measured by the GRAVITY Collaboration provides significantly tighter constraints on the model parameters than using astrometric and radial velocity data alone. In the case of the Yukawa-like black hole, this allows us to constrain the scale length parameter $\lambda$ at the 1$\sigma$ confidence level.

Future advances in astrometric and spectroscopic techniques, together with extended monitoring of short-period S stars such as S4711, S62, and S4714 \citep{peissker2020s62}, as well as continued GRAVITY interferometric observations of the S2 star \citep{fernandez2023constraining}, are expected to place stronger constraints on the black hole solutions. Such data will provide deeper insight into the nature of Sgr\,A* and offer a reliable tool to test alternative solutions of black holes and theories of gravity beyond GR.

\section*{Acknowledgements}
This research was partially funded by Universidad Nacional Aut\'onoma de M\'exico (UNAM) through Direcci\'on General de Asuntos del Personal Acad\'emico (DGAPA) grants IA101123 and IN110522. CN gratefully acknowledges financial support by Secretar\'ia de Ciencia, Humanidades, Tecnolog\'ia e Innovaci\'on (SECIHTI) through a graduate fellowship. ACO acknowledges the SECIHTI programme Ciencia B\'asica y de Frontera 2023-2024, project CBF2023-2024-1102. We thank the anonymous Referee for their constructive comments on the Manuscript. Numerical simulations were performed at \textsc{Laboratorio de Modelos y Datos} (LAMOD) at Instituto de Ciencias Nucleares, UNAM.

\section*{Data Availability}
A machine-readable version of the data utilized in this study is publicly accessible in the online version of \cite{gillessen2017update}.

\bibliographystyle{mnras}
\bibliography{aeireferences}

\appendix

\section{Equations of geodesic motion}\label{sec:EulerLagrange}
\let\origaddcontentsline\addcontentsline
\def\addcontentsline#1#2#3{\origaddcontentsline{#1}{#2}{#3}\let\addcontentsline\origaddcontentsline}

\let\addcontentsline\origaddcontentsline

Here we provide details on the Lagrangian formalism to obtain the equations of motion of a test particle moving on a given spherical space-time described by a metric tensor $g_{\alpha\beta}$.
We consider timelike geodesics parametrized by their proper time $\tau$, and described by the set of coordinates $(x^\alpha(\tau)) = (t(\tau),r(\tau),\theta(\tau),\phi(\tau))$.
We use a Lagrangian in geometrized units for a free particle moving along such a timelike geodesic,
\begin{eqnarray}
\mathcal{L}=\frac{1}{2}g_{\alpha \beta }\frac{\mathrm{d} x^{\alpha }}{\mathrm{d} \tau }\frac{\mathrm{d} x^{\beta }}{\mathrm{d} \tau } = -\frac{1}{2}.
\end{eqnarray}
The Euler-Lagrange equations are then invoked to obtain second-order equations of motion,
\begin{eqnarray}
\frac{\mathrm{d} }{\mathrm{d} \tau }\left (\frac{\partial \mathcal{L}}{\partial \dot{x}^{\gamma }}  \right )-\frac{\partial \mathcal{L}}{\partial x^{\gamma }}=0,
    \label{eq:EcsL}
\end{eqnarray}
where an overdot represents differentiation with respect to $\tau$. Given initial conditions, we integrate numerically the set of equations using the python library \textsc{scipy} \citep{2020SciPy-NMeth}.

Since all of the considered space-times have the Killing vector fields $\bm{\xi}_t = \partial/\partial t$ and $\bm{\xi}_\phi = \partial/\partial \phi$, both specific energy ($\mathcal{E}$) and specific angular momentum ($\ell$) are conserved along timelike geodesics, and are calculated as 
\begin{eqnarray}
\mathcal{E} &=&-\frac{\partial \mathcal{L}}{\partial \dot{t}} ,  \\
\ell &=&\frac{\partial \mathcal{L}}{\partial \dot{\phi}}.   
\label{eq:ELConserved}
\end{eqnarray}

\section{Code validation}
\label{sec:CodeVal}
The following crucial factors were taken into account in order to evaluate the robustness of the code used to obtain the posterior density distributions.

\begin{figure}
\begin{center}
\includegraphics[width=0.5\textwidth]{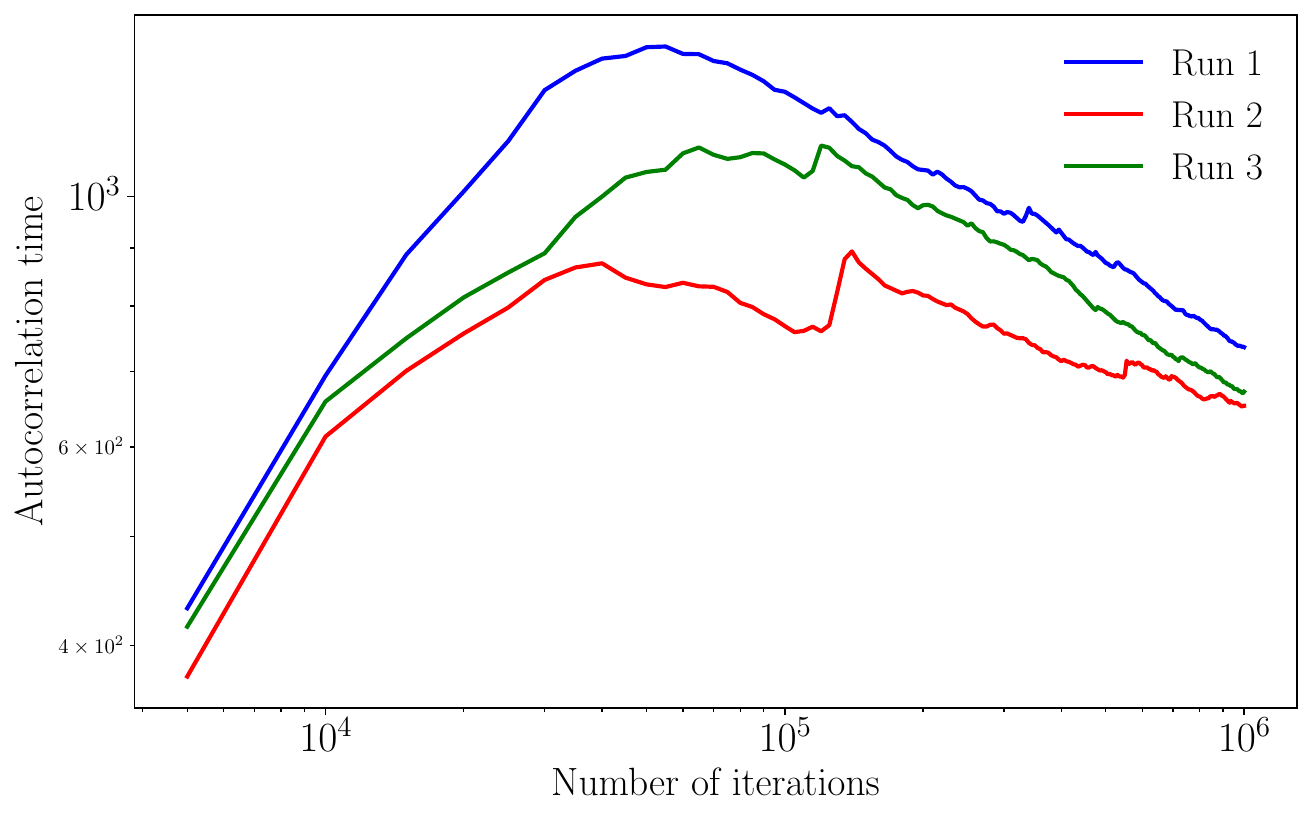}
\caption{Representative example of the autocorrelation time as a function of the number of iterations, for three runs of the MCMC method under the same initial conditions and parameters, for the Schwarzschild model.}
\label{fig:ConvTest}
\end{center}
\end{figure}

The integrated autocorrelation time ($\tau_{f}$) was used to diagnose the convergence of the parameter distributions, which is the recommended criterion when working with the \textsc{emcee} package \citep{foreman2013emcee}. The integrated autocorrelation time was stored for different runs of the MCMC method, with the different space-time models, in a range of one million iterations. The results in general show that this value saturates after a certain number of iterations in this range, and in all cases the number of iterations exceeds 100 $\tau_{f}$. A representative example of this is shown in Fig. \ref{fig:ConvTest}. Similarly, the mean acceptance fraction was monitored to ensure that it was within an acceptable range, since this indicator should be in the range of 0.2-0.5 \citep{foreman2013emcee}. Also, a burn-in period was considered at the beginning of each MCMC run.

Additionally, in our MCMC runs, the parameter proposals are generated from a combination of moves, which allows for a more efficient sampling in a high-dimensional model. In particular, we use a weighted mixture of \textit{StretchMove}, \textit{DEMove}, and \textit{DESnookerMove}, which are implemented in the \textsc{emcee} Python package.

Finally, to corroborate the implementation of the Python library \textsc{dynesty} \citep{speagle2020dynesty, koposov2023joshspeagle}, we use summary (run) plots to verify convergence to the maximum likelihood estimate. The code was executed for $500,000$ iterations in each case, and log-evidence convergence was also verified.

\section{Robustness of the parameter estimates to the likelihood weighting}
\label{sec:likelihood_sensitivity}
In our analysis, the total likelihood is written as the sum of astrometric, spectroscopic, and precession contributions.  Since the precession constraint itself is derived from a combination of pre- and post-pericenter astrometry, the separation of these terms introduces an implicit assumption of independence.  Several studies of stellar orbits around the Galactic Centre \citep[e.g.][]{della2022orbital} adopt a more conservative approach in which all likelihood terms are uniformly downweighted by a factor of $\sqrt{2}$, thereby ensuring that no information is counted twice in the global fit.

To assess the robustness of our results with respect to this potential source of over-weighting, we repeated the Schwarzschild parameter inference using the same sampler configuration as in the main text but multiplying all likelihood contributions by $1/\sqrt{2}$.  This provides a direct test of the sensitivity of the inferred orbital and dynamical parameters to the chosen likelihood structure.

Table~\ref{table:LikelihoodComparison} summarizes the posterior medians and $1\sigma$ credible intervals obtained with the standard likelihood (Column~A) and with the uniformly downweighted likelihood (Column~B). All parameters remain fully consistent within uncertainties.  The central values shift by amounts well below the $1\sigma$ level, while the posterior widths increase only modestly, as expected from the reduced statistical weight.

\begin{table}
\centering
\begin{tabular}{lcc}
\hline
\multicolumn{1}{c}{Parameter} &
\multicolumn{1}{c}{A: Standard likelihood} &
\multicolumn{1}{c}{B: $\sqrt{2}$-downweighted} \\
\hline

\addlinespace[5pt] 
$r_{d}$ (pc) & $8,115^{+220}_{-217}$ & $8,148^{+308}_{-302}$ \\

\addlinespace[5pt]
$M$ ($10^6 \mathrm{M}_{\odot}$) & $4.236^{+0.231}_{-0.217}$ & $4.268^{+0.331}_{-0.303}$ \\

\addlinespace[5pt]
$a$ (mas) & $127.20^{+1.35}_{-1.27}$ & $127.03^{+1.88}_{-1.75}$ \\

\addlinespace[5pt]
$e$ & $0.88480^{+0.00274}_{-0.00270}$ & $0.88442^{+0.00383}_{-0.00375}$ \\

\addlinespace[5pt]
$t_{\mathrm{a}} - 2010$ (yr) & $0.3972^{+0.0137}_{-0.0135}$ & $0.3978^{+0.0193}_{-0.0191}$ \\

\addlinespace[5pt]
$i$ ($^{\circ}$) & $133.80^{+0.51}_{-0.51}$ & $133.87^{+0.71}_{-0.72}$ \\

\addlinespace[5pt]
$\omega$ ($^{\circ}$) & $65.26^{+0.70}_{-0.70}$ & $65.17^{+0.99}_{-0.98}$ \\

\addlinespace[5pt]
$\Omega$ ($^{\circ}$) & $226.06^{+0.71}_{-0.70}$ & $225.98^{+1.00}_{-0.99}$ \\

\addlinespace[5pt]
$x_0$ (mas) & $0.14^{+0.44}_{-0.44}$ & $0.14^{+0.62}_{-0.62}$ \\

\addlinespace[5pt]
$y_0$ (mas) & $-2.10^{+0.68}_{-0.69}$ & $-2.03^{+0.96}_{-0.96}$ \\

\addlinespace[5pt]
$v_{x, 0}$ (mas yr$^{-1}$) & $0.124^{+0.046}_{-0.046}$ & $0.124^{+0.065}_{-0.065}$ \\

\addlinespace[5pt]
$v_{y, 0}$ (mas yr$^{-1}$) & $0.012^{+0.072}_{-0.072}$ & $0.006^{+0.101}_{-0.102}$ \\

\addlinespace[5pt]
$v_{z, 0}$ (km s$^{-1}$) & $24.4^{+8.0}_{-7.9}$ & $25.1^{+11.4}_{-11.3}$ \\

\addlinespace[5pt]
\hline
\end{tabular}
\caption{Posterior median values and $1\sigma$ credible intervals for the Schwarzschild model obtained using the standard likelihood (Column~A) and a uniformly downweighted likelihood in which all terms are multiplied by $1/\sqrt{2}$ (Column~B).  The two analyses yield consistent results, indicating that the parameter estimates in the main text are robust to the treatment of the precession constraint.}
\label{table:LikelihoodComparison}
\end{table}

\section{MCMC Corner plots}\label{sec:Corn}
In this appendix, we present corner plots for the seven space-times considered in this work. The posterior distributions show confidence levels at 1$\sigma$, 2$\sigma$, and 3$\sigma$ of the parameter set in each model. The limits at 1$\sigma$ depicted in the figures correspond to those presented in Table \ref{table:BestFit}.
 
 \begin{figure*}
  \begin{center}
   \includegraphics[width=1\textwidth]{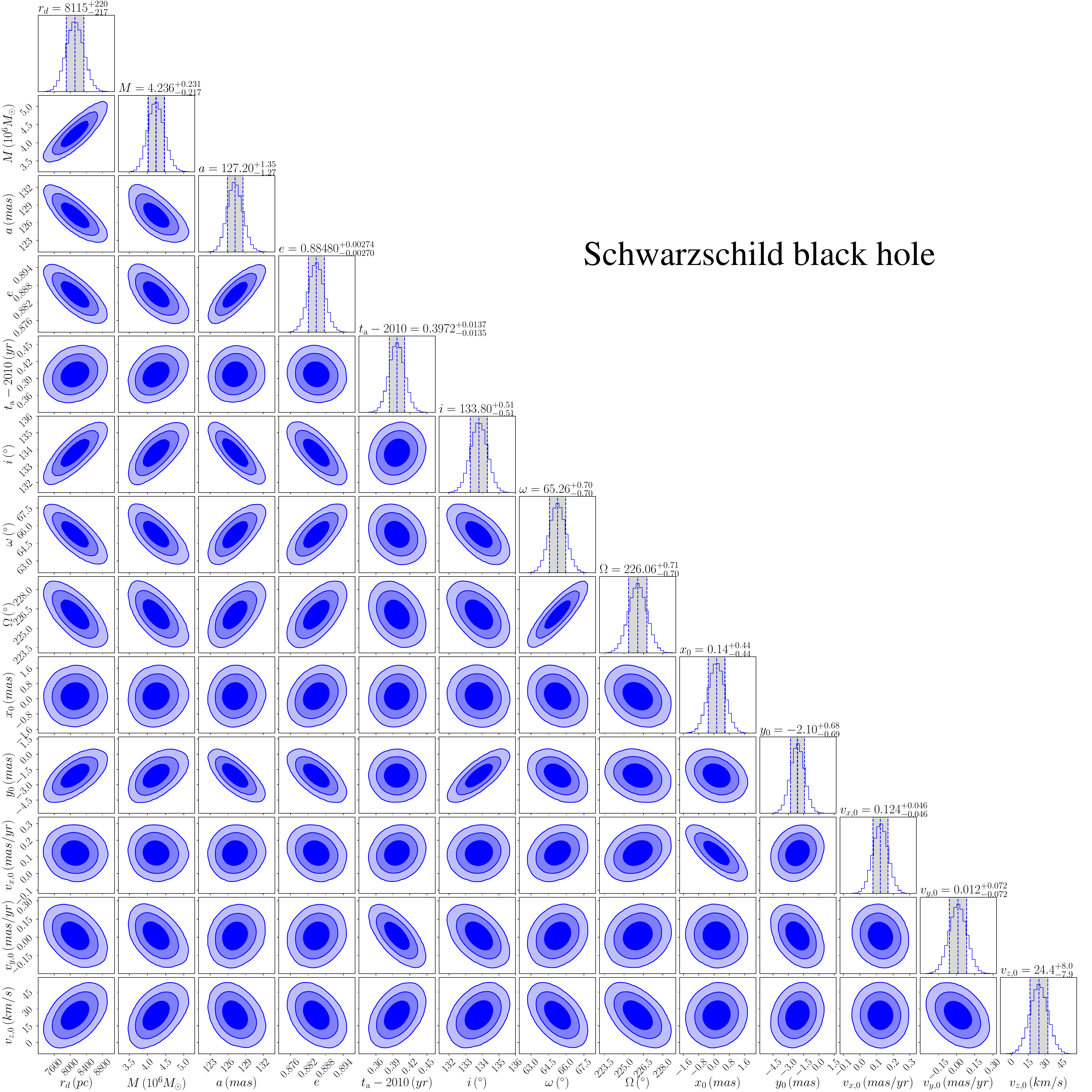}
   \caption{Posterior distribution contours with 68\% (1$\sigma$), 95\% (2$\sigma$), and 99.7\% (3$\sigma$) confidence levels for the parameters that model the orbit of the S2 star in Schwarzschild space-time. The gray-shaded region in the histograms corresponds to 1$\sigma$ confidence interval. The mean value and the 1$\sigma$ range of the corresponding parameter are shown above each column, and are represented in the histogram by the three vertical lines.}
   \label{fig:SchCorner}
  \end{center}
 \end{figure*}

  \begin{figure*}
  \begin{center}
   \includegraphics[width=1\textwidth]{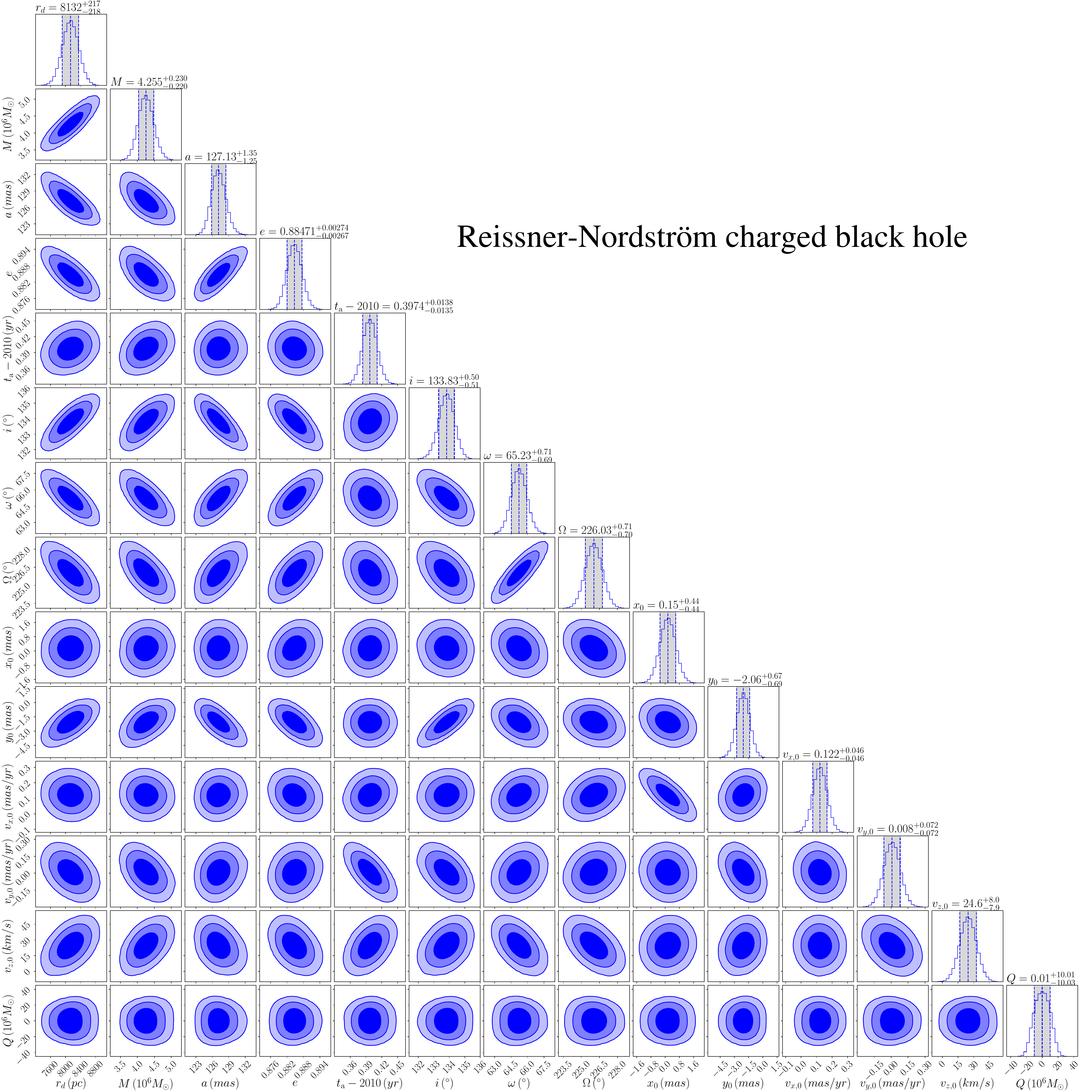}
   \caption{Posterior distribution contours with 68\% (1$\sigma$), 95\% (2$\sigma$), and 99.7\% (3$\sigma$) confidence levels for the parameters that model the orbit of the S2 star in Reissner-Nordstr\"om space-time. The gray-shaded region in the histograms corresponds to 1$\sigma$ confidence interval. The mean value and the 1$\sigma$ range of the corresponding parameter are shown above each column, and are represented in the histogram by the three vertical lines.}
   \label{fig:RN_Corner}
  \end{center}
 \end{figure*}

    \begin{figure*}
  \begin{center}
   \includegraphics[width=1\textwidth]{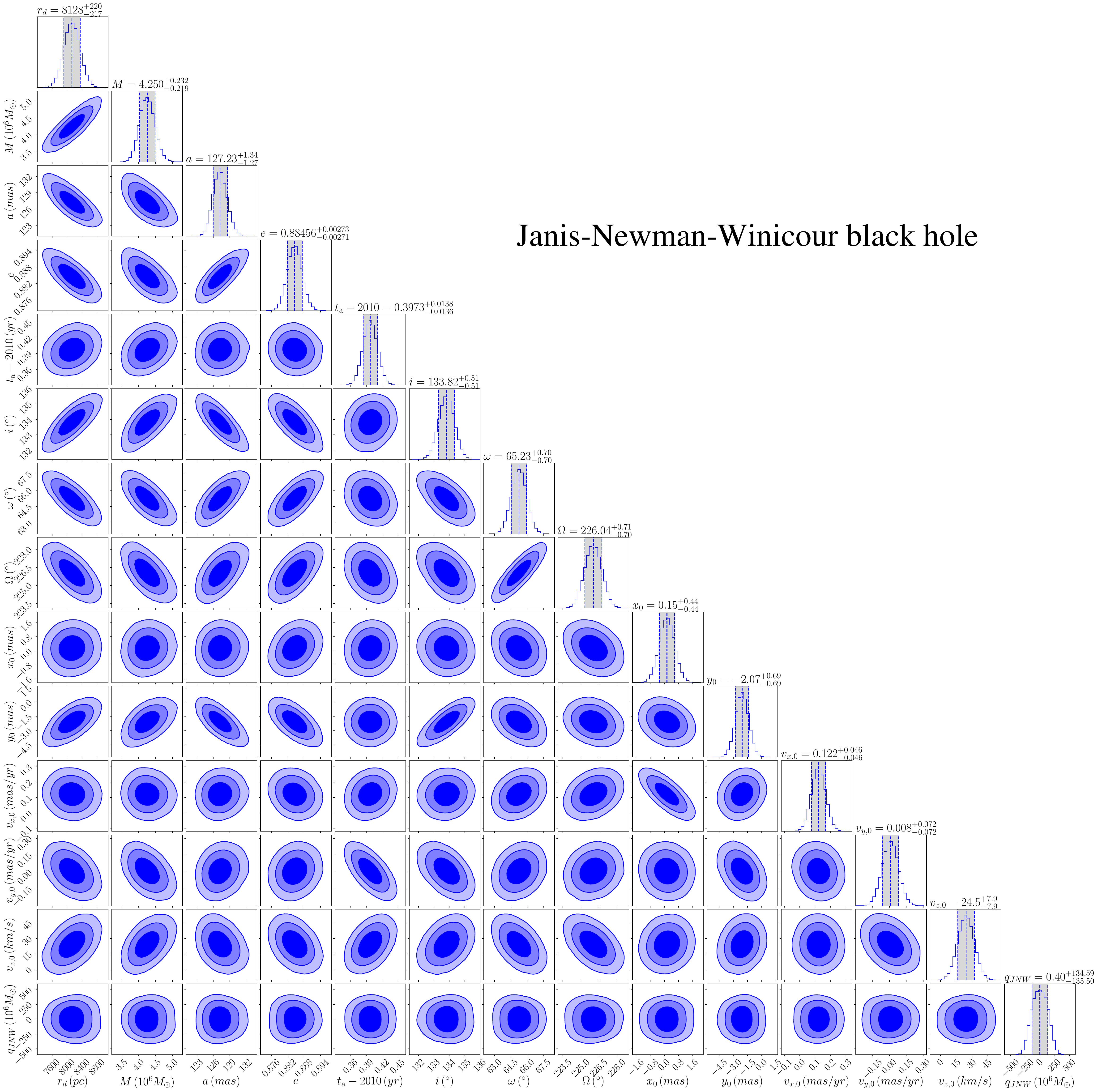}
   \caption{Posterior distribution contours with 68\% (1$\sigma$), 95\% (2$\sigma$), and 99.7\% (3$\sigma$) confidence levels for the parameters that model the orbit of the S2 star in Janis-Newman-Winicour space-time. The gray-shaded region in the histograms corresponds to 1$\sigma$ confidence interval. The mean value and the 1$\sigma$ range of the corresponding parameter are shown above each column, and are represented in the histogram by the three vertical lines.}
   \label{fig:JNW_Corner}
  \end{center}
 \end{figure*}

   \begin{figure*}
  \begin{center}
   \includegraphics[width=1\textwidth]{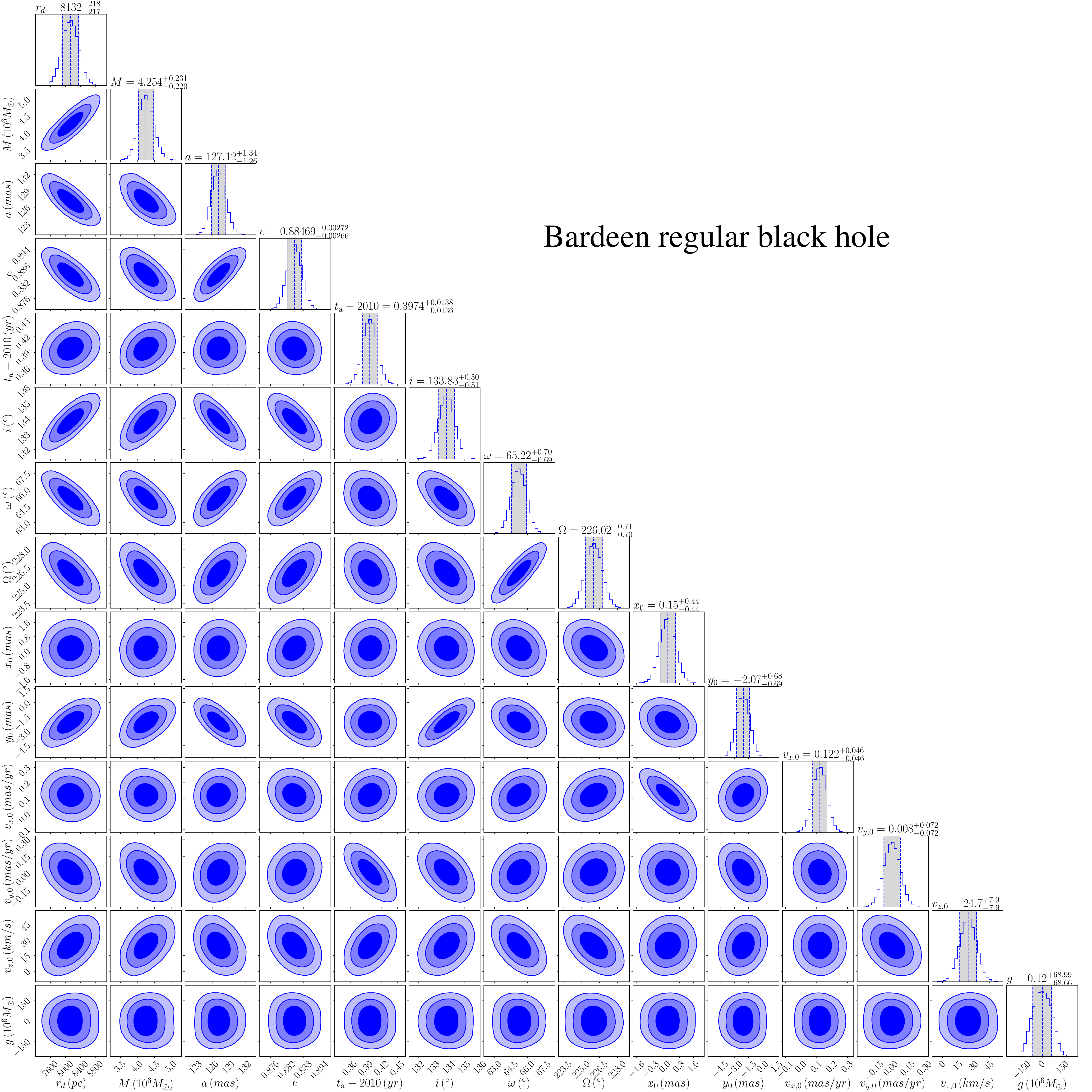}
   \caption{Posterior distribution contours with 68\% (1$\sigma$), 95\% (2$\sigma$), and 99.7\% (3$\sigma$) confidence levels for the parameters that model the orbit of the S2 star in Bardeen space-time. The gray-shaded region in the histograms corresponds to 1$\sigma$ confidence interval. The mean value and the 1$\sigma$ range of the corresponding parameter are shown above each column, and are represented in the histogram by the three vertical lines.}
   \label{fig:Bardeen_Corner}
  \end{center}
 \end{figure*}

\begin{figure*}
  \begin{center}
   \includegraphics[width=1\textwidth]{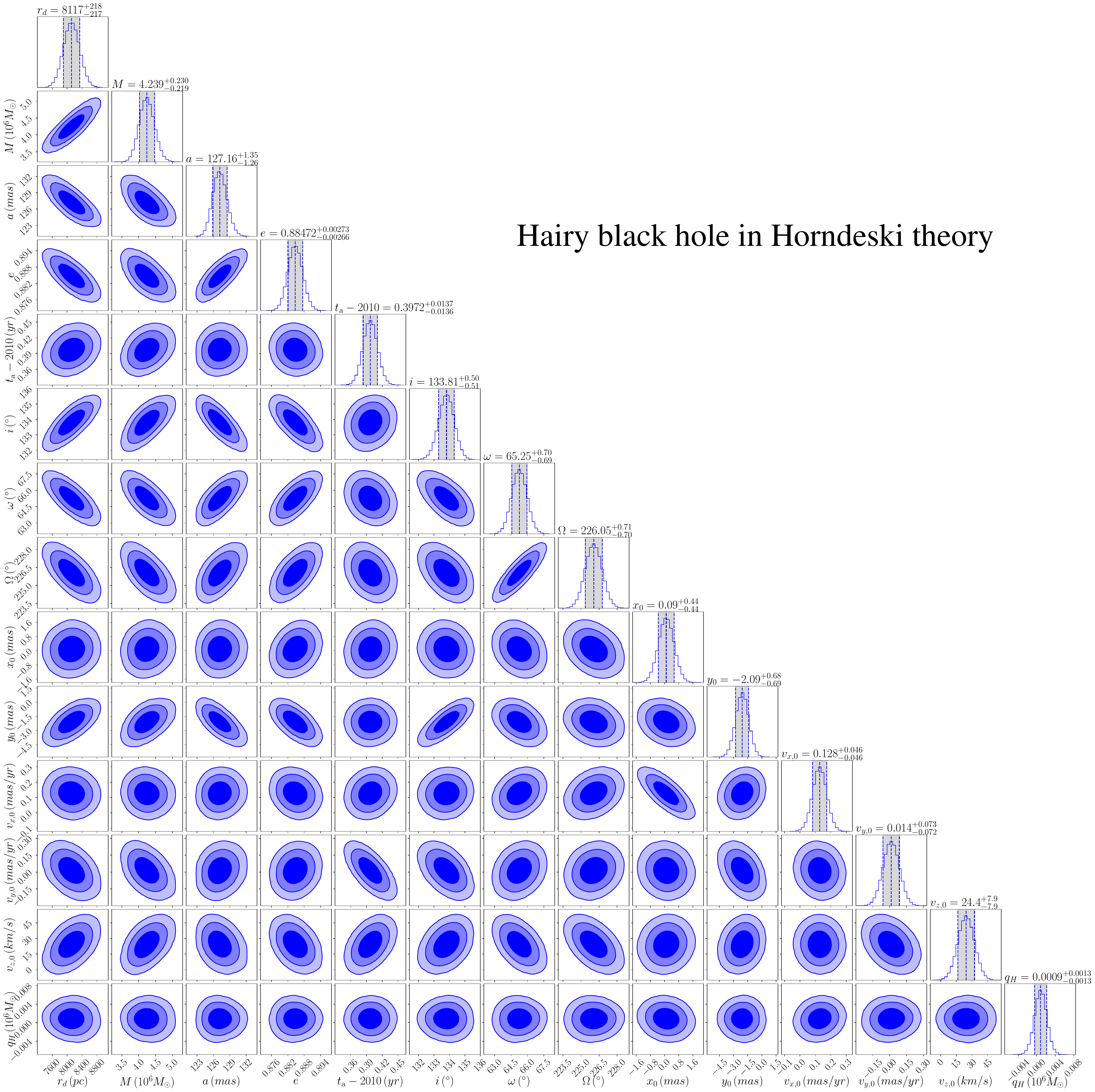}
   \caption{Posterior distribution contours with 68\% (1$\sigma$), 95\% (2$\sigma$), and 99.7\% (3$\sigma$) confidence levels for the parameters that model the orbit of the S2 star in Horndeski space-time. The gray-shaded region in the histograms corresponds to 1$\sigma$ confidence interval. The mean value and the 1$\sigma$ range of the corresponding parameter are shown above each column, and are represented in the histogram by the three vertical lines.}
   \label{fig:Horndeski_Corner}
  \end{center}
 \end{figure*}

 \begin{figure*}
  \begin{center}
   \includegraphics[width=1\textwidth]{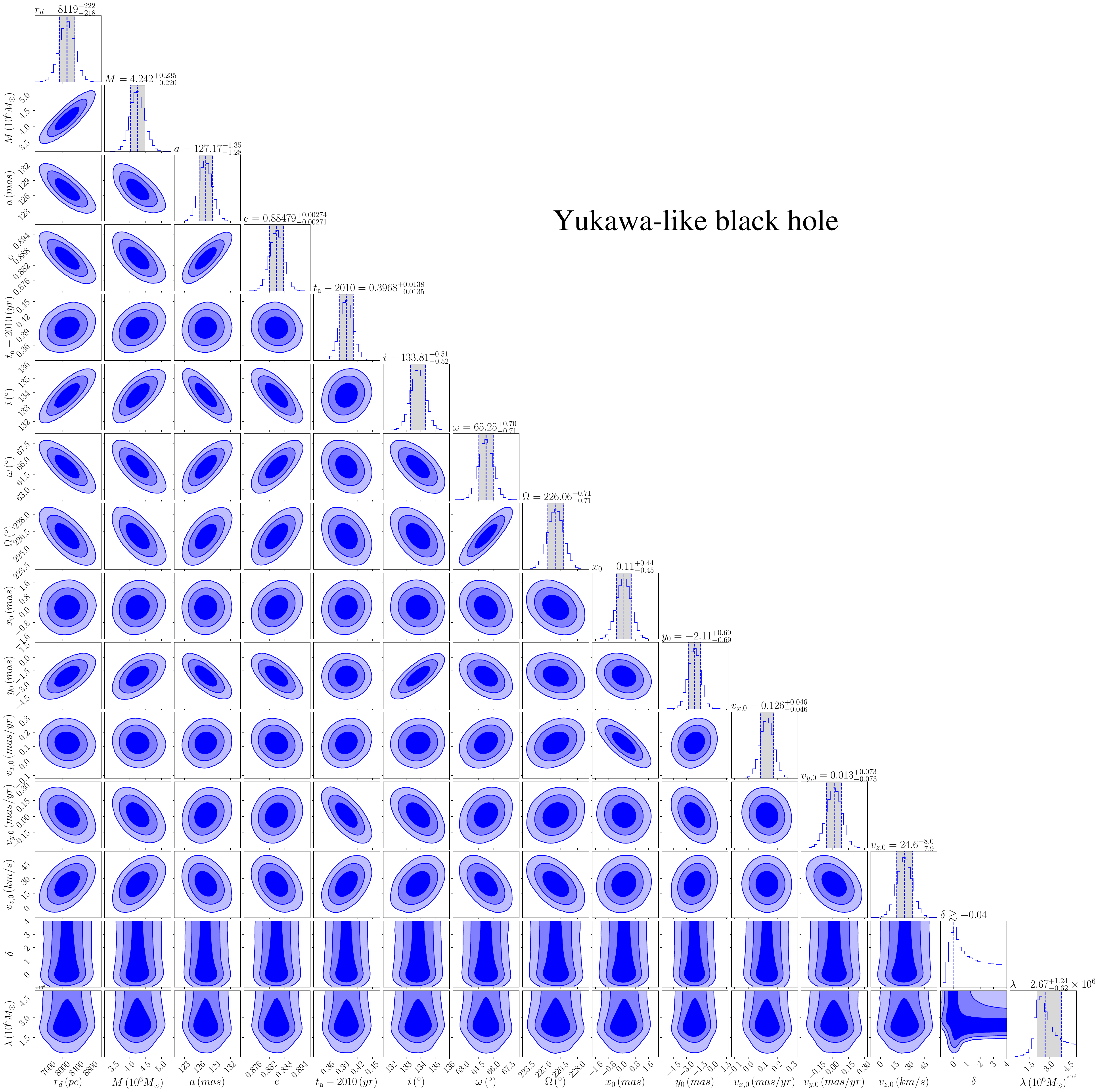}
   \caption{Posterior distribution contours with 68\% (1$\sigma$), 95\% (2$\sigma$), and 99.7\% (3$\sigma$) confidence levels for the parameters that model the orbit of the S2 star in Yukawa-like black hole. The gray-shaded region in the histograms corresponds to 1$\sigma$ confidence interval. The mean value and the 1$\sigma$ range of the corresponding parameter are shown above each column, and are represented in the histogram by the three vertical lines. For the strength of the Yukawa-like potential, the lower bound obtained at 1$\sigma$ is shown.}
   \label{fig:Yukawa_corner}
  \end{center}
 \end{figure*}

 \begin{figure*}
  \begin{center}
   \includegraphics[width=1\textwidth]{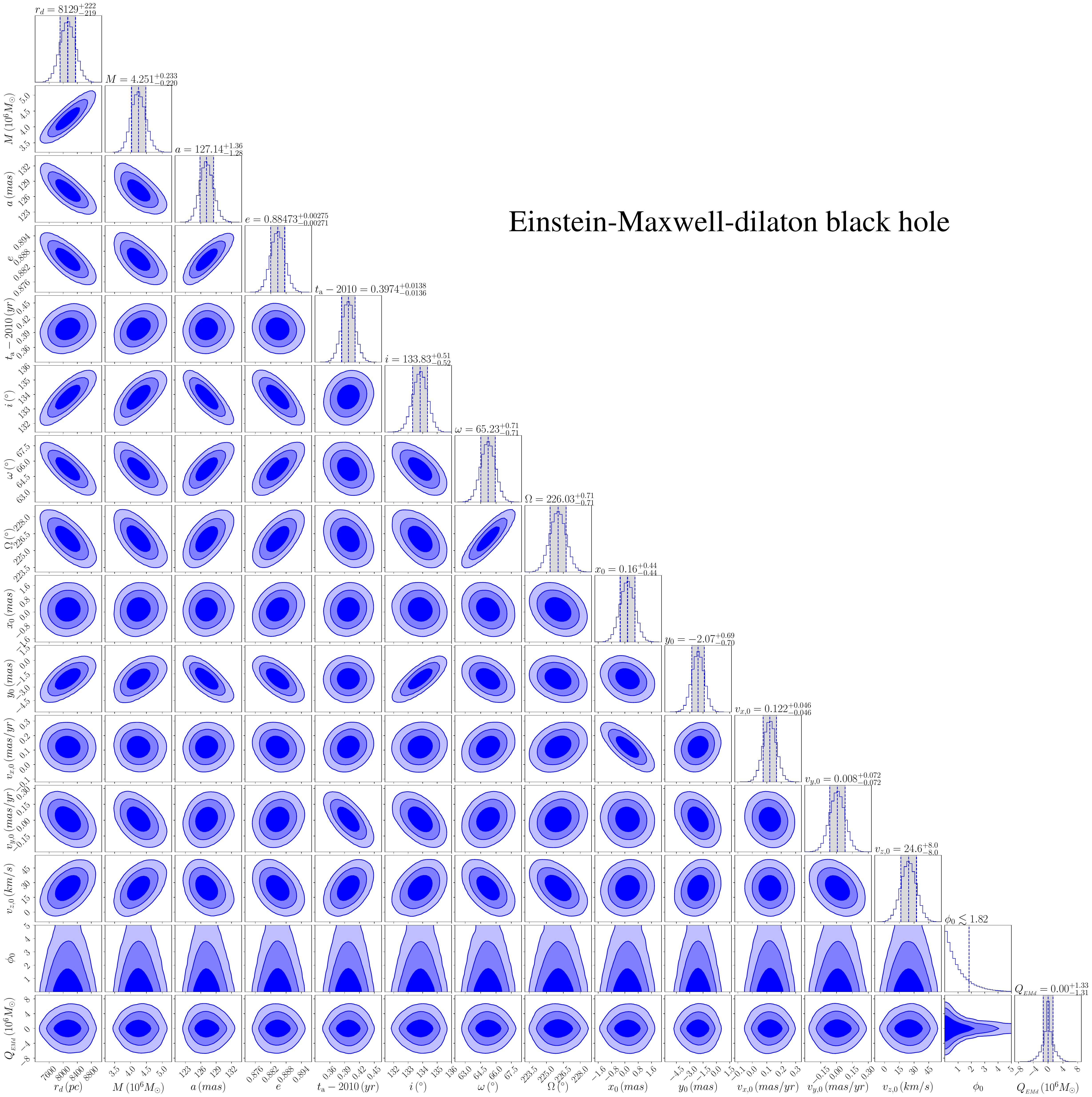}
   \caption{Posterior distribution contours with 68\% (1$\sigma$), 95\% (2$\sigma$), and 99.7\% (3$\sigma$) confidence levels for the parameters that model the orbit of the S2 star in EMd space-time. The gray-shaded region in the histograms corresponds to 1$\sigma$ confidence interval. The mean value and the 1$\sigma$ range of the corresponding parameter are shown above each column, and are represented in the histogram by the three vertical lines. For the asymptotic value of the dilatonic field, the upper limit obtained at 1$\sigma$ is shown.}
   \label{fig:EMD_Corner}
  \end{center}
 \end{figure*}

\label{lastpage}
\end{document}